\documentclass{cernrep}
\usepackage{color}
\usepackage[flushleft]{threeparttable}
\usepackage{easyReview}
\usepackage{longtable}
\usepackage{subcaption}
\usepackage[unicode=true,pdfusetitle,
 bookmarks=true,bookmarksnumbered=false,bookmarksopen=false,
 breaklinks=false,pdfborder={0 0 1},backref=false,colorlinks=true,linkcolor=blue,linktoc=page]
 {hyperref}
\makeatletter
\usepackage{mciteplus}
\usepackage{hyperref}

\newcommand\footnoteref[1]{\protected@xdef\@thefnmark{\ref{#1}}\@footnotemark}
\makeatother
\usepackage{ifthen} % for conditional statements
\newboolean{articletitles}
\setboolean{articletitles}{true} % False removes titles in references

\usepackage{titlepage}
\addauthor{ECFA Higgs / Top / Electroweak Factory Study}{}

\title{Focus topics for the ECFA study on Higgs / Top / EW factories}

\abstract{In order to stimulate new engagement and trigger some concrete studies in areas where further work would be beneficial towards fully understanding the physics potential of an $e^+e^-$ Higgs / Top / Electroweak factory, we propose to define a set of focus topics. The general reasoning and the proposed topics are described in this document.
}

\begin{document}

\titlepage

\tableofcontents
\clearpage
\section*{Motivation}%~\label{sec:motivation}
\addcontentsline{toc}{section}{Motivation}

\begin{center}
{\itshape
  Coordinators:
  Jorge~de~Blas,
  Patrick Koppenburg,
  Jenny List,
  Fabio~Maltoni
}
\end{center}

In the last update of the European Strategy for Particle Physics~\cite{CERN-ESU-015}, an $e^+e^-$ collider serving as a Higgs factory was identified as the highest-priority next collider. The
US P5 panel also endorses an off-shore Higgs factory, located
in either Europe or Japan, to advance studies of the Higgs boson following the HL-LHC~\cite{P5-2023}.

The ECFA Higgs / Top / Electroweak Factory study~\cite{ECFA-WHF-indico} has been set up to build an $e^+e^-$ community, bringing people together across the various $e^+e^-$ projects to share expertise and tools and to work coherently on scientific and technical topics.

In this document we propose specific areas in which the ECFA study could reach significantly beyond the state-of-the-art understanding of the physics potential of future $e^+e^-$ Higgs / top / EW factories. The proposed topics do not aim to comprehensively map the physics program of a future Higgs factory. Instead, they should serve to:

\begin{itemize}
\item complete the current overall picture where (most) necessary 
\item give guidance to people who would like to contribute to the ECFA study
\item highlight processes particularly suitable to study the interplay of the three working areas of physics potential, analysis methods and detector performance.
\end{itemize}

The topics can therefore act as a vehicle for new engagement and collaboration.  They are intended as a basis that could be expanded later. The initiative should build on existing analysis tools and samples that can be shared among the projects and developed cooperatively. We therefore try to point out where existing examples, including analysis code and datasets, could be taken as a starting point, particularly by new entrants. All experimental simulation studies are strongly encouraged to use the {\sc Key4hep} framework~\cite{Key4hep:2021zms,Key4hep:2023xhe}. This will translate into new tools usable by the whole community and thoroughly tested, and will improve already existing or interfaced tools.

While the proposed focus topics are considered as high priority targets, the final report of the ECFA study is clearly expected to give a broader picture of the physics potential of the future Higgs / Top / Electroweak Factory, summarising all relevant results. 
Many studies have already contributed to this picture, but new ideas are still emerging. Those interested in collaborating on subjects going beyond the focus topics presented in this document are welcome to contact the conveners of the relevant group directly.

The list of topics is presented in Table~\ref{t:sqrt_s_topic} indicating also the responsible WG1 sub-group and the relevant centre-of-mass energies. The following sections give high-level summaries of the state-of-the-art and open questions for each topic.

\section*{General references and MC samples}
\addcontentsline{toc}{section}{General references and MC samples}

Here we assemble for the convenience of the reader a number of general references on the major Higgs / Top / EW Factories:
\begin{enumerate}
\item General state-of-the-field reports: 
\begin{itemize}
  \item the Briefing Book for the last European Strategy Update~\cite{EuropeanStrategyforParticlePhysicsPreparatoryGroup:2019qin}
  \item the Snowmass Energy Frontier~\cite{Narain:2022qud} and overall summary~\cite{Butler:2023glv}
\end{itemize}
\item Recent Workshops:
\begin{itemize}
  \item LCWS2023~\cite{LCWS2023}
  \item FCC Week 2023~\cite{FCC2023}
  \item CEPC Workshop 2023~\cite{CEPC2023}
  \item C$^3$ Workshop 2023~\cite{CCC2023}
  \item 2nd ECFA Workshop on Higgs / Top / EW Factories~\cite{ECFA2023}
\end{itemize}  
\item Project Design Reports
\begin{itemize}
  \item FCC-ee volume of the FCC CDR~\cite{FCC:2018evy}
  \item CEPC CDR\cite{CEPCStudyGroup:2018rmc, CEPCStudyGroup:2018ghi}
  \item CLIC CDR, Updated Staging Baseline, and 2018 Summary Report~\cite{Linssen:2012hp, CLIC:2016zwp, CLICdp:2018cto}
  \item ILC TDR~\cite{ILCInternationalDevelopmentTeam:2022izu, Behnke:2013xla, ILC:2013jhg, Adolphsen:2013jya, Adolphsen:2013kya, Behnke:2013lya}
\end{itemize}  
\item Detector Concepts
\begin{itemize}
  \item The ILD Interim Report~\cite{ILDConceptGroup:2020sfq}
  \item A recent description and update plans of SiD~\cite{Breidenbach:2021sdo}
  \item Description of the CLIC detector~\cite{CLICdp:2018vnx}
  \item Description of CLD~\cite{Bacchetta:2019fmz}
  \item Description of IDEA~\cite{Antonello:2020tzq}
  \item A recent description of the CEPC detector~\cite{CEPCPhysics-DetectorStudyGroup:2019wir}
\end{itemize}
\end{enumerate}

For many processes, large scale MC samples exist, either publicly available, or, for the latest version of fully simulated detector concepts, via the respective detector concept group. A summary for general samples is listed here; more specific information will be given for each focus topic if applicable. These existing samples should be sufficient as a starting point, while typically dedicated needs will arise as the work on each topic progresses.
We stress that for every topic the goals include having multiple generators capable of providing Monte Carlo of comparable precision available, interfaced to {\sc Key4hep}.

\begin{itemize}
      \item Publicly available events at generator level (with ILC beam spectrum) can be found at \url{http://ilcsnowmass.org/} 
      \item For access to ILD full simulation samples, please contact Ties Behnke (ties.behnke@desy.de cc ild-et@desy.de)
      \item For access to SiD full simulation samples, please contact Andrew White (awhite@uta.edu) and Marcel Stanitzki (marcel.stanitzki@desy.de) 
      \item For access to CLICdp samples please contact Andr\'e Sailer (andre.philippe.sailer@cern.ch)
      \item For access to CLD samples please contact Andr\'e Sailer (andre.philippe.sailer@cern.ch) 
      \item For access to IDEA samples please contact Patrizia Azzi (patrizia.azzi@cern.ch) and Emmanuel Perez (Emmanuel.Perez@cern.ch)

   \end{itemize}

\begin{table*}[t]
  \centering
    \caption{Overview of focus topics and relevant centre-of-mass energies. 
    Energies applicable to the considered topic are indicated with '\checkmark'.
      \label{t:sqrt_s_topic}
    }
  \vspace{4pt}
    {
     \begin{threeparttable}
     \begin{tabular}{rl |c | c c c c c} 
        %\ctoprule
        %\crowcolor
        \hline
        &
        &
        &
        \multicolumn{5}{c}{Relevant $\sqrt{s}$  [GeV]} 
        \\
        \multicolumn{2}{l|}{Topic} &
        Lead group &
        $91$
         &
        $161$ &

        $240$--$250$ &

        $350$--$380$  &

        $\ge 500$  % \tnote{a}   % (AFZ) see below
        \\
        \hline
%-----------------------------------
        \phantom{0}\ref{sec:HtoSS} & {\bfseries} HtoSS  &
           HTE &
           & % MZ 
           & % 161 GeV 
           \checkmark & % 240/250 GeV
           \checkmark & % 350 / 365 / 380 GeV
           \checkmark % \ge 500 GeV
          \\ [0.1cm]
%-----------------------------------
        \phantom{0}\ref{sec:ZHang} & {\bfseries} ZHang   &
           HTE (GLOB)&
           & % MZ 
           & % 161 GeV 
           \checkmark & % 240/250 GeV
           \checkmark  &% 350 / 365  / 380GeV
           \checkmark % \ge 500 GeV
          \\ [0.1cm]
%-----------------------------------
        \phantom{0}\ref{sec:Hself} & {\bfseries} Hself   &
           GLOB &
           & % MZ 
           & % 161 GeV 
           \checkmark & % 240/250 GeV
           \checkmark & % 350 / 365  / 380GeV
           \checkmark % \ge 500 GeV
          \\ [0.1cm]
%-----------------------------------
       \phantom{0}\ref{sec:Wmass} & {\bfseries} Wmass    &
           PREC &
           & % MZ 
           \checkmark & % 161 GeV 
           \checkmark & % 240/250 GeV
           \checkmark & % 350 / 365 / 380 GeV
            \checkmark % \ge 500 GeV
          \\ [0.1cm]
%-----------------------------------
        \phantom{0}\ref{sec:WWdiff} & {\bfseries} WWdiff &
           GLOB &
           & % MZ 
           & % 161 GeV 
           \checkmark & % 240/250 GeV
           \checkmark & % 350 / 365 / 380 GeV
           \checkmark  % \ge 500 GeV
          \\ [0.1cm]
%-----------------------------------
        \phantom{0}\ref{sec:TTthres} & {\bfseries} TTthres  &
           GLOB (HTE) &
           & % MZ 
           & % 161 GeV 
           & % 240/250 GeV
           \checkmark  & % 350 /365 / 380 GeV
            \checkmark % \ge 500 GeV
          \\ [0.1cm]
%-----------------------------------
        \phantom{0}\ref{sec:LUMI} & {\bfseries} LUMI  &
           PREC &
           \checkmark & % MZ 
           \checkmark & % 161 GeV 
           \checkmark & % 240/250 GeV
           \checkmark & % 350 / 365 / 380 GeV
           \checkmark % \ge 500 GeV
          \\ [0.1cm]
%-----------------------------------
        \phantom{0}\ref{sec:EXscalar} & {\bfseries} EXscalar  &
           SRCH &
            & % MZ 
            & % 161 GeV 
           \checkmark & % 240/250 GeV
           \checkmark & % 350 / 365 / 380 GeV
           \checkmark % \ge 500 GeV
          \\ [0.1cm]
%-----------------------------------
        \phantom{0}\ref{sec:LPPs} & {\bfseries} LLPs   &
           SRCH &
           \checkmark & % MZ 
           \checkmark & % 161 GeV 
           \checkmark & % 240/250 GeV
           \checkmark & % 350 / 365 / 380 GeV
           \checkmark % \ge 500 GeV
          \\ [0.1cm]
%-----------------------------------
        \ref{sec:EXtt} & {\bfseries} EXtt   &
           SRCH &
            & % MZ 
            & % 161 GeV 
            & % 240/250 GeV
           \checkmark &  % 350 / 365  / 380 GeV
           \checkmark % \ge 500 GeV
          \\ [0.1cm]
%-----------------------------------
        \ref{sec:CKMWW} & {\bfseries} CKMWW   &
          FLAV &
             & % MZ 
           \checkmark &% 161 GeV 
           \checkmark &% 240/250 GeV
           \checkmark & % 350 / 365 / 380 GeV
           \checkmark  %\ge 500 GeV
          \\ [0.1cm]
%-----------------------------------
        \ref{sec:BKtautau} & {\bfseries} BKtautau  &
           FLAV &
           \checkmark & % MZ 
             &% 161 GeV 
             &% 240/250 GeV
             & % 350 /365 / 380GeV
             %\ge 500 GeV
          \\ [0.1cm]
%-----------------------------------
        \ref{sec:TwoF} & {\bfseries} TwoF  &
          HTE (PREC) &
           \checkmark & % MZ 
           \checkmark & % 161 GeV 
           \checkmark & % 240/250 GeV
           \checkmark & % 350 /365 /380GeV
           \checkmark %\ge 500 GeV
          \\ [0.1cm]
%-----------------------------------
        \ref{sec:BCfrag} & {\bfseries} BCfrag and Gsplit &
           PREC (FLAV) &
           \checkmark & % MZ 
           \checkmark & % 161 GeV 
           \checkmark & % 240/250 GeV
           \checkmark & % 350 /365 /380GeV
           \checkmark %\ge 500 GeV
          \\ [0.1cm]
%-----------------------------------
       % 15 {\bfseries} Gsplit  &
        %   PREC (FLAV) &
        %   \checkmark & % MZ 
        %   \checkmark & % 161 GeV 
        %   \checkmark & % 240/250 GeV
        %   \checkmark & % 350 /365 /380GeV
        %   \checkmark %\ge 500 GeV
        %  \\ [0.1cm]
        \hline
      \end{tabular}
     \end{threeparttable}   
     }
\end{table*}

\subsection*{Contact \& Further Information}
\begin{itemize}
\item Gitlab wiki:\\ \url{https://gitlab.in2p3.fr/ecfa-study/ECFA-HiggsTopEW-Factories/-/wikis/}
\item Email the ECFA Higgs factory WG1 coordinators:\\ \url{mailto:ECFA-Workshop-Higgs-factory-coords@cern.ch}
\item or email the conveners of the groups listed in the sections below.
\end{itemize}

\clearpage
\section{{\bfseries HtoSS} --- $e^+e^- \to Zh$:  $h \to s\bar{s}$  ($\sqrt{s}=240/250$\,GeV)}
\label{sec:HtoSS}
% \section{{\bfseries HtoSS} -- $e^+e^- \to Zh$:  $h \to ss$  ($\sqrt{s}=240/250$\,GeV) (JdB)}
\begin{center}\itshape
Expert Team: John Alison, Matthew Basso, Valentina Cairo, Valerio Dao, Loukas Gouskos, Karsten K\"oneke, Yotam Soreq, Taikan Suehara, Caterina Vernieri
\end{center}

The core of the physics program at Higgs factories is the determination of the absolute Higgs-strahlung ($Vh$) cross-section with the least possible model-dependence along with precision measurements of the Higgs boson couplings to fermions. Both aspects have been studied thoroughly within the last European Strategy Update~\cite{deBlas:2019rxi} and more recently in the context of the 2021 Snowmass community exercise~\cite{Dawson:2022zbb,Narain:2022qud}. The study of the coupling of the Higgs boson to the light quarks, i.e. up, down, and strange quarks was considered nearly impossible due to the small branching ratio when assuming SM couplings as well as the difficulty in identifying the flavour of quark-initiated jets (flavour tagging). Nevertheless, the exploration of the Higgs coupling to the strange quark $y_s$ has emerged with increasing interest, given also its tight connections with detector technologies and layout optimisation.
Enabling sensitivity to inclusive $h\rightarrow s\bar{s}$ would allow for a complete exploration of the second-generation Yukawa couplings, and go beyond the current LHC limited reach for $y_s$ via $h\to\phi\gamma$ decays~\cite{Kagan:2014ila,ATLASZandHToPhiOrRho}.

At the LHC, in addition to the small branching fraction, the rare  $h\rightarrow s\bar{s}$ decay mode is made inaccessible by the current detector capabilities. In fact, one of the most powerful handles to identify a strange-quark-initiated jet (strange-tagging) is the possibility to distinguish between kaons and pions up to tens of GeV in momentum. This relies on dedicated detector subsystems which are not included in the LHC multi-purpose detectors. Furthermore, the overwhelming multi-jet production rate at the LHC inhibits the study of strange, up, and down quark couplings with inclusive $h\rightarrow q\bar{q}$ decays, in addition to the dominant $h\rightarrow b\bar{b}$ decay mode. Therefore, future Higgs factories present a unique opportunity to probe new physics frontiers with the strange quark and to design detectors that enable such studies.

Along with the SM scenario, this signature provides the possibility of testing several Beyond the Standard Model (BSM) theories that allow for extended Higgs sectors. Models with additional Higgs doublets have new Yukawa couplings which need not be directly proportional to the SM fermion masses. For a discussion of the possible mechanisms leading to modified Yukawa couplings, see Refs.~\cite{Cerri:2018ypt,Cepeda:2019klc}.

As an example, a second class of BSM models exists where the 125\,GeV Higgs couples predominantly to the third generation~\cite{Ghosh:2015gpa,Altmannshofer:2015esa,2HDM}. This results in very different decay branching ratios of the additional heavy Higgs bosons ($H$). The largest production mode of the neutral Higgs bosons would be from a $c\bar{c}$ initial state, while the charged Higgs bosons would be predominantly produced from a $c\bar{s}$ initial state. The most interesting decay modes include $H/A \rightarrow c\bar{c}$, $t\bar{c}$, $\mu\bar{\mu}$~\cite{ATLASHAmm, CMSAmm}, and $\tau\mu$~\cite{ATLASHmutauetau, CMSHmutauetau} and $H^\pm \rightarrow c\bar{b}$, $c\bar{s}$~\cite{CMSHcs2020}, and $\mu\nu$.

Another example, models exhibiting spontaneous flavour violation (SFV)~\cite{Egana-Ugrinovic:2018znw} would allow for new Yukawa couplings either to the up or the down quarks with no relation to the quark masses. 
A two Higgs doublet model with up-type SFV could thus have large couplings to the $d$ and $s$ quarks, and the new Higgs states would be produced in quark fusion, with decays to gauge and Higgs bosons and quarks~\cite{Egana-Ugrinovic:2019, Egana-Ugrinovic:2021}. 
If the observed 125\,GeV Higgs boson is an admixture of a SM-like Higgs and one of the new Higgs states, its couplings to the first- or second-generation quarks can be significantly larger than predicted in the SM, leading to large deviations in the Higgs boson branching ratios.

In the past years, preliminary proof-of-concept investigations at future colliders have been performed. Some of them~\cite{Bedeschi:2022rnj} focus primarily on strange-tagging algorithms and some others~\cite{Duarte-Campderros:2018ouv,albert2022strange} include also their application to $h\rightarrow s\bar{s}$ searches, interpretations in BSM frameworks, and potential detector designs. The assumptions and the detector concepts used in these studies differ. For example, a fast simulation approach targeting the IDEA detector concept at the FCC was used in Ref.~\cite{Bedeschi:2022rnj}, while the results presented in Ref.~\cite{albert2022strange} rely on full simulation samples of the ILD detector concept at the ILC, but assume truth-based Particle Identification (PID) information. These choices were primarily driven by readiness at the time the studies were performed. Nevertheless, all the results show promising avenues and motivate more in-depth explorations and future harmonisation. In order to advance the field, the ECFA {\bfseries HtoSS} expert team has identified possible directions, which are listed below.

\subsection*{Theoretical, phenomenological and MC generator targets}
Expanding the BSM interpretations of the studies that have already been performed or developing new simulation-based analyses targeting specific BSM scenarios would enlarge the physics case for strange tagging at future colliders. In particular, we welcome studies in the following areas:
\begin{itemize}
    \item Detailed understanding of how to extract the Higgs-strange coupling strength from a BR($h \to s\bar{s}$) measurement, given contributions from Dalitz decays, e.g, $h \to g^{\ast}(\to s\bar{s})g$ or $h \to \gamma^{\ast}(\to s\bar{s})\gamma$.
    \item BSM models predicting deviations in $h \to s\bar{s}$, e.g., SUSY or composite Higgs --- see Refs.~\cite{Cepeda:2019klc,Cerri:2018ypt};
    \item BSM models predicting, for example, charged Higgs bosons with large branching ratios in final states including strange quarks, e.g., 2HDM $H^{+} \to cs$ BR $\approx 50\%$;
    \item $s \bar s$ vs.\ $b \bar b$ in BSM models: gain from $s \bar s$;
    % \item $h \to bs$?
    \item BSM flavour structure and $h\to s\bar{s}$ signal.
\end{itemize}

\subsection*{Target physics observables}
Several physics quantities will be investigated:
\begin{itemize}
    \item $e^+e^- \to Zh$ with $h \to ss$ ($Z \to $ anything) at $\sqrt{s}=240/250$\,GeV (this has been the only target so far, but it will be relevant to explore also higher centre-of-mass energies, which, in turn, enable different Higgs production modes);
    \item projected precision on the branching fraction and the differential cross-section in $\cos{\theta_s}$;
    \item flavour-changing decays are very rare in the SM, for example, BR($h \to bs$) $\simeq 10^{-7}$. New physics models, which can be encapsulated by an EFT, allow larger values. % This will come for free in a multi-flavour analysis (already ongoing)?
\end{itemize}

\subsection*{Target analysis techniques}
The performed proof-of-concept studies \cite{albert2022strange, Bedeschi:2022rnj} showed that to improve the results there will be a large need for more powerful background rejection techniques as well as a potentially more global approach in the extraction of the Higgs couplings. Two areas of particular interest will be:
\begin{itemize}
    \item diboson background suppression;
    \item signal extraction (fit discriminant variables, counting experiments, etc.).
\end{itemize}

\subsection*{Target methods to be developed}
In collaboration with the Reconstruction and Detector groups, the impact from the following features will have to be evaluated when estimating the analysis sensitivity reach, including:
\begin{itemize}
    \item control of strange-tagging related systematic uncertainties;
    \item reconstruction of in-flight decays, e.g., $K^0_{\rm S} \to \pi^+\pi^-$;
    \item strangeness-tagging with ML techniques and compared with anti-$b$-tagging techniques;
    \item $s$ vs $\bar{s}$ separation;
    \item complementarity of particle identification (ID) techniques for charged hadrons in momentum reach (from $dN/dx$, $dE/dx$, ToF, RICH);
    \item understanding the contribution from $g \to s\bar s$ (from single jets) to strange-tagging performance and analysis sensitivity.
\end{itemize}

\subsection*{Target detector performance aspects}
The obtained results will inform the community on two crucial aspects:
\begin{itemize}
    \item dependence of the precision on physics observables on particle ID, strange-tagging, and reconstruction capabilities;
    \item technology benchmarks for sub-detectors.
\end{itemize}  

\subsection*{Generation and Simulation needs}
Full simulation samples will be needed to perform the studies listed above. Samples for $e^+e^- \to f \bar{f} h$ at $\sqrt{s}=240/250$\,GeV and $350/380/550$\,GeV are available as indicated in the general samples listed in the motivation. In the years to come, it will be important to iterate with simulation experts on $s\bar{s}$ correlations and fragmentation uncertainties in order to account for more realistic systematic uncertainties.

\subsection*{Existing tools / examples}
% \textcolor{red}{Valentina: I would maybe remove this which is a bit out of data and instead add the references to the papers we have? which are already cited above}
% Mention that there are tools (both for FCC and ILC) available

There are several existing tools and analysis codes available. At the time of writing, this includes examples for ILC and FCC-ee. However, due to ongoing developments, in case you would like to get actively engaged, please contact us directly (see below), such that we can point you to the up-to-date tools and code repositories. 

\subsection*{Contact \& Further Information}
\begin{itemize}
\item Gitlab wiki: \url{https://gitlab.in2p3.fr/ecfa-study/ECFA-HiggsTopEW-Factories/-/wikis/FocusTopics/HtoSS}
\item Sign up for egroup: ECFA-WHF-FT-HtoSS@cern.ch via \url{http://simba3.web.cern.ch/simba3/SelfSubscription.aspx?groupName=ecfa-whf-ft-HtoSS}
\item and/or email the coordinators of this ECFA WG1 focus topic:\\ \url{mailto:ECFA-WHF-FT-HtoSS-coordinators@cern.ch}
\end{itemize}

\clearpage
\section{{\bfseries ZHang} --- $Zh$ angular distributions and CP studies}
\label{sec:ZHang}
%\section{{\bfseries ZHang} -- $e^+e^- \to ZH$: reconstruction of production and decay angles   ($\sqrt{s}=240/250$ GeV)}
\begin{center}
{\itshape 
Expert Team: Ivanka Bozovic, Chris Hays, Markus Klute, Sandra Kortner, Cheng Li, Ken Mimasu, Gudrid Moortgat-Pick 
}
\end{center}

Angular distributions in $Zh$ production can be used to increase sensitivity to both CP-even and CP-odd interactions of the Higgs boson.  The Higgs self-coupling vertex appears at next-to-leading order in $Zh$ production, and a global analysis of CP-even interactions including angular distributions from this process can improve the sensitivity to the self-coupling.  The presence of a CP-odd component in Higgs-boson interactions can be probed by reconstructing the Higgs and $Z$ boson decay planes, or by measuring and utilising the polarisations of the Higgs-boson decay particles.  These CP-odd interactions could provide an ingredient to explain the observed matter-antimatter asymmetry in the universe.  Prior analyses of $Zh$ production have found good sensitivity to CP-odd interactions, and a further understanding of this sensitivity is a primary goal of this topic.

\subsection*{CP-odd interactions}

Analyses involving the coupling of the observed Higgs state, $h$, and two electroweak vector bosons, 
$V=Z,W$, only offer access to the CP-odd coupling through loop effects, since there is no direct renormalisable coupling 
between a pseudoscalar and a pair of gauge bosons.  Couplings of the observed Higgs boson to fermions are therefore important, since the CP-even and CP-odd components can have a similar magnitude.  Experimental evidence for CP-violating couplings would clearly point to BSM physics, and interpretations in terms of both effective couplings and concrete models are necessary.

\subsubsection*{Current status and challenges}
The Snowmass 2021 study~\cite{Dawson:2022zbb,Gritsan:2022php} quantified the sensitivity of various processes to CP-odd interactions in terms of the fraction of the CP-odd amplitude relative to the total, $f_{\rm CP}$.  For $hVV$ couplings, the fraction is derived from the following amplitude parametrisation~\cite{Anderson:2013afp}:
\begin{equation}
    A(hV_1V_2)= \frac{1}{v}\Big[ a_1^{hVV} m_{V_1}^2 \epsilon^*_{V_1}\epsilon^*_{V_2} + a_2^{hVV} {f^*}^{(1)}_{\mu\nu}{f^*}^{(2),\mu\nu} + \frac{1}{2}a_3^{hVV} \epsilon^{\mu\nu\rho\sigma}{f^*}^{(1)}_{\mu\nu}{f^*}^{(2)}_{\rho\sigma}\Big],
\end{equation}
\noindent
where $f^{(i),\mu\nu} = \epsilon^{\mu}_i q_i^{\nu} - \epsilon^{\nu}_i q_i^{\mu}$ is the field strength tensor of a gauge boson with momentum $q_i$ and polarisation vector $\epsilon_i$, and $\tilde{f}^{(i),\mu\nu} = 1/2 \epsilon^{\mu\nu\alpha\beta}f_{\alpha\beta}$ is the conjugate field strength tensor. Using this parametrisation the CP-odd fraction is defined as
\begin{equation}
    f^{hVV}_{\rm CP} = \frac{|a_3^{hVV}|^2}{\sum_i |a_i^{hVV}|^2(\sigma_i/\sigma_3)}.
\end{equation}

\noindent
The Snowmass-defined target for CP-sensitive measurements for $hVV$-processes is $f_{\rm CP}<10^{-5}$, whereas for the $hff$ and loop-induced $h\gamma\gamma$, $h\gamma Z$, and $hgg$ processes it is about $f_{\rm CP}<10^{-2}$.  The target is based on a benchmark model point of the two-Higgs-doublet model~\cite{Shu:2013uua} that provides sufficient CP violation to provide the observed baryon asymmetry in the universe.  The chosen parameters, including $\tan\beta=1$, provide a CP mixing angle of 0.1.  For other model parameters, such as $\tan\beta <1$, the mixing angle can be reduced by a factor of two or more, indicating that it is well motivated to aim for sensitivities below the Snowmass target value.

Direct access to CP-odd couplings requires CP-odd observables, and those involving triple product correlations have been extensively explored~\cite{Rindani:2010pi}. In addition to the momenta, the polarisation of the initial or final fermions can be used, including longitudinally as well as transversely polarised beams.
The Snowmass study updated the CP-analysis in the $Zh$-process, finding a 68\% C.L. sensitivity to $f^{hVV}_{\rm CP}$ of $3.9~(2.9) \times 10^{-5}$ at a centre-of-mass energy of 250~(350)~GeV.    At an energy of 1000~(500) GeV, the sensitivity improves to $3~(13) \times 10^{-6}$ with 10~(5)~ab$^{-1}$ of integrated luminosity.  The study used leptonic $Z$-boson decays and assumed equivalent results for hadronic decays, while employing a simplified detector model and neglecting background.  
At an energy of 1 TeV, the $ZZ$-fusion process also contributes, and a recent ILC study using a full ILD simulation including background found a sensitivity to $f^{hVV}_{\rm CP}$ of $16 \times 10^{-6}$ for this process with 8~ab$^{-1}$ of integrated luminosity. % PoS(EPS-HEP2023)404, reference to be added
For the HL-LHC the Snowmass study quoted the CMS projected  sensitivity of $2.5 \times 10^{-6}$ using the VBF production process and assuming $a_3^{WW}=\cos^2\theta_W a_3^{ZZ}$. 

The Higgs-to-fermion coupling is challenging for CP studies.  The theoretical target of $f_{\rm CP}<10^{-2}$ should be possible for the $htt$-coupling at either HL-LHC or a 1~TeV $e^+e^-$ collider. Better sensitivity is possible for the $h\tau\tau$ coupling, for which the target is achievable at the HL-LHC and at the $240-250$ GeV $e^+e^-$ collider with nominal luminosity. 
The Snowmass study found a sensitivity of $f_{\rm CP}=0.01$ for the $h\tau\tau$ coupling at $\sqrt{s}=250$~GeV and at $\sqrt{s}=350$~GeV.  
The anticipated HL-LHC sensitivity is $f_{\rm CP}=0.07$ for the $h\tau\tau$ interaction, along with 0.12 and 0.24 for the $hgg$ and $htt$ vertices, respectively.

\subsubsection*{Potential studies}
The sensitivity of an $e^+e^-$ collider to the CP structure of several Higgs interactions has been established.  Further studies can determine whether there is scope to improve the sensitivity, or to extend it to additional interactions.  Possibilities include:
\begin{itemize}
\item a complete implementation of the $Zh$ analysis including all $Z$ decays and backgrounds into an existing experimental framework; 
\item using angular information or an optimal observable to improve sensitivity to the CP structure of the $hZZ$ vertex;
\item a joint constraint on the CP-even and CP-odd components of the $hZZ$ vertex using pseudo-observables or the SM effective field theory (SMEFT), rather than just the CP-odd fraction; 
\item analysis of the CP sensitivity to the $hZZ$ vertex in an asymmetric collider, as in the HALHF design~\cite{Foster:2023bmq}; and 
\item improvements in sensitivity from exploiting beam polarisation.
\end{itemize}

\noindent
In addition to the sensitivity studies, an expanded interpretation framework connecting the SMEFT to specific model scenarios could be used to clarify the coverage of an $e^+e^-$ collider to the CP-odd interaction strengths that can explain the baryon asymmetry in the universe.

Further studies beyond the angular distributions of $Zh$ production are also possible.  A high priority is the complete implementation of the most sensitive channel $h\to \tau\tau$ into an experimental setup, including systematic uncertainties and potential improvements of the detailed polarisation reconstruction.  Other possibilities include an investigation into the sensitivity of other processes, such as VBF Higgs production at high $\sqrt{s}$, or other interactions, such as  $hZ\gamma$ and $h\gamma\gamma$ (which require an internal or external photon conversion).  Updates to the projected HL-LHC sensitivity would provide more accurate comparisons to the $e^+e^-$ collider studies.

These studies will likely require development of reconstruction techniques, including: tracking improvements for $\tau$-lepton and quark-jet reconstruction, combining both timing and spacepoint information; improvements in jet-charge tagging to separate the quark and anti-quark directions; improvements in vertexing and particle identification required to reconstruct the $\tau$-lepton polarisation.  Additional Monte Carlo samples may be required for various CP scenarios, and for various polarisations.

\subsection*{CP-even interactions}

Sensitivity of the $e^+e^- \to Zh$ cross section to the Higgs self-coupling, $\lambda_{hhh}$ has been established at NLO~\cite{McCullough:2013rea}.  An indirect precision of $\approx 25\%$ on this coupling is possible through precise measurements of $ZH$ production at $\sqrt{s} = 240$~GeV~\cite{Dawson:2022zbb}.  The inclusion of angular information may improve this sensitivity. 

The expected HL-LHC precision of 50\% for the predicted SM Higgs self-coupling is a factor of two higher.  Current searches at the LHC constrain the ratio ($\kappa_\lambda=\lambda_{hhh}/\lambda_{hhh}^{\rm SM}$) of the observed self-coupling to the Standard Model prediction to be in the range $-0.4$ ($-1.2$) to $6.3$ ($6.5$) using Run 2 ATLAS~\cite{ATLAS:2022jtk} (CMS~\cite{CMS:2022dwd}) data, assuming that only this coupling is affected by new processes in single-Higgs and di-Higgs production.  The HL-LHC projections could be updated in light of the final Run 2 results.

Angular asymmetry observables in $Zh$ production have been shown to constrain three independent combinations of dimension-6 SMEFT coefficients at lowest order~\cite{Beneke:2014sba,Craig:2015wwr}.
In the global context, these were found to be of limited importance in disentangling Higgs couplings in an analysis that focused on bosonic and Yukawa-like Higgs couplings~\cite{Durieux:2017rsg}, provided that sufficient other inputs from $W^+W^-$ production and VBF were included.  This was also observed in a more general context in Ref.~\cite{DeBlas:2019qco}, where it is argued that a combination of runs at different energies (notably the $Z$-pole) would already be sufficient to lift the flat directions that the angular information would remove. The latter analysis also made use of statistically optimal observables, finding modest gains in sensitivity.  The inclusion of polarisation information has been found to provide additional improvements~\cite{Barklow_2018}.

\subsubsection*{Potential studies}

 The existing studies can be extended to a global SMEFT analysis to determine the generality of the $e^+ e^- \to Zh$ process sensitivity to the Higgs self-coupling.  Several specific activities are possible:
\begin{itemize}
    \item determine whether angular or other observables can target the sensitivity to the self-coupling, possibly in conjunction with different centre-of-mass energies and beam polarisations; 
    \item perform a complete NLO analysis of the $ZH$ process within the context of a global SMEFT analysis, including constraints from other measurements; and 
    \item extend the global SMEFT analysis to dimension-8 operators and all terms at order $1/\Lambda^4$.
\end{itemize}

\noindent
The extension of the SMEFT analysis to order $1/\Lambda^4$ is particularly valuable given that both CP-odd and CP-even operators contribute to many observables at this order.  The optimisation of the analysis to SMEFT interactions can include the use of quantum information observables, which have been shown to provide sensitivity at order $1/\Lambda^4$ for diboson processes~\cite{Aoude:2023hxv}.  Finally, concrete models such as those motivated by baryogenesis can be used to validate the SMEFT analysis and to directly compare the sensitivity of various observables to these models.

\subsection*{Contact \& Further Information}
\begin{itemize}
\item Gitlab wiki: \url{https://gitlab.in2p3.fr/ecfa-study/ECFA-HiggsTopEW-Factories/-/wikis/FocusTopics/ZHang}
\item Sign up for egroup: ECFA-WHF-FT-ZHANG@cern.ch via \url{http://simba3.web.cern.ch/simba3/SelfSubscription.aspx?groupName=ecfa-whf-ft-zhang}
\item and/or email the conveners of ECFA WG1 HTE group:\\ \url{mailto:ecfa-whf-wg1-hte-conveners@cern.ch}
\end{itemize}

\clearpage
\section{{\bfseries Hself} --- Determination of the Higgs self-coupling}
\label{sec:Hself}
% mainfile: ../benchmarks

%\section{{\bfseries Hself} -- Determination of the Higgs self-coupling (JdB)}
\newcommand{\htb}[1]{\textcolor{blue}{#1}}
\newcommand{\htr}[1]{\textcolor{red}{#1}}
\newcommand{\lahhh}{\lambda_{hhh}}
\newcommand{\lahhH}{\lambda_{hhH}}
\newcommand{\laSM}{\lambda_{hhh}^{\rm SM}}
\newcommand{\kala}{\kappa_\lambda}
\begin{center}
%{\itshape Expert Team: Junping Tian, Gauthier Durieux, Ricardo Goncalo, Sven Heinemeyer, Michael Peskin, Philipp Roloff, Roberto Salerno}
{\itshape Expert Team: Gauthier Durieux, Ricardo Gon\c{c}alo, Sven Heinemeyer, Michael Peskin, Philipp Roloff, Roberto Salerno, Junping Tian }
\end{center}

To measure the Higgs self-coupling, $\lahhh$, at future $e^+e^-$ colliders, the classic approach relies on double Higgs production processes which requires $\sqrt{s}\ge500$ GeV for the $e^+e^-\to Zhh$ process or above 1 TeV for the $e^+e^-\to \nu\bar{\nu}hh$ process. 
%The diagrams and cross sections are shown in Fig.~\ref{}. 
This approach has been explored by ILC and CLIC using full detector simulations ~\cite{Durig:2016jrs,Roloff:2019crr}. The standard projections, which assume the SM value of $\lahhh$, indicate that $\lahhh$ can be measured with a precision of $\sim 27\%$ at the ILC500, and $10\%$ is reachable at the ILC1000 and CLIC3000. The HL-LHC projection is about $50\%$. The challenge in the Higgs pair production approach mainly comes from small cross sections of signal processes, $O(0.1)$\,fb, and the analysis techniques to effectively suppress background which is many orders of magnitude higher. Moreover, the predicted values of Higgs pair production cross sections result from a strong interference of diagrams with and without $\lahhh$.
There is also a conceptual challenge that how we can determine $\lahhh$ based on double Higgs production cross sections if other unknown couplings from the BSM enter 
%\JB{I'd say only BSM. In the SM, we know what couplings enter here and only $\lahhh$ is experimentally not tested. Thus, if the couplings are unknown, they cannot be SM.} 
in the same processes, such as the quartic coupling $hhZZ$ or contact interaction $eeZh$. This question has been addressed in Refs.~\cite{Barklow:2017awn, DiVita:2017vrr} and it turns out that other unknown couplings will not have a significant impact to the precision of $\lahhh$ based on global fits in the SMEFT, thanks to the tree-level contribution from $\lahhh$ in the double-Higgs cross section. 
%\htr{SH: what about $\lahhH$, which can have a strong impact?}

There is another emerging approach which probes the quantum effects by the Higgs self-coupling in the single-Higgs process~\cite{McCullough:2013rea}. This is now actively pursued by all $e^+e^-$ collider projects, in particular FCC-ee and CEPC. This approach is enabled by the prospect that the single-Higgs production cross section can be measured to a precision well below $1\%$, which reaches the level of the quantum effect from $\lahhh$, for instance at $\sqrt{s}=240$~GeV $\delta\sigma_{Zh}=2\delta \kappa_Z+1.7\%\cdot\delta\kappa_\lambda$, with $\kappa_\lambda=\lahhh/\laSM$. In this approach we need to control all the systematic uncertainties and theory uncertainties to well below $1\%$. The biggest challenge comes from the fact that there are potentially many more parameters (e.g. the above $\kappa_Z$ and others) that contribute to the cross section~\cite{DiVita:2017vrr}, in particular at one-loop level, and may induce uncertainties at level of $1\%$. Thus the conceptual question is whether it is possible at all to measure $\lahhh$,
or this measurement should rather be seen as a consistency test of the SM?
%or it only provides a test to SM? 
Observables at two different $\sqrt{s}$ have been shown to be helpful in lifting the degeneracy with $\kappa_Z$~\cite{DiVita:2017vrr}. The NLO contributions from top-quark in single-Higgs process are very relevant in terms of degeneracy with $\lahhh$ as studied in Refs.~\cite{Durieux:2018ggn,Jung:2020uzh}, and the dedicated constraints on top-quark EW couplings from HL-LHC or direct top pair production at future $e^+e^-$ colliders will play a very important role in the precision of Higgs couplings including $\lahhh$. 
%\htr{SH: again it is unclear what happens if $\lahhH$ enters.}

\begin{figure}[h]
	\centering
	\includegraphics[width=\textwidth]{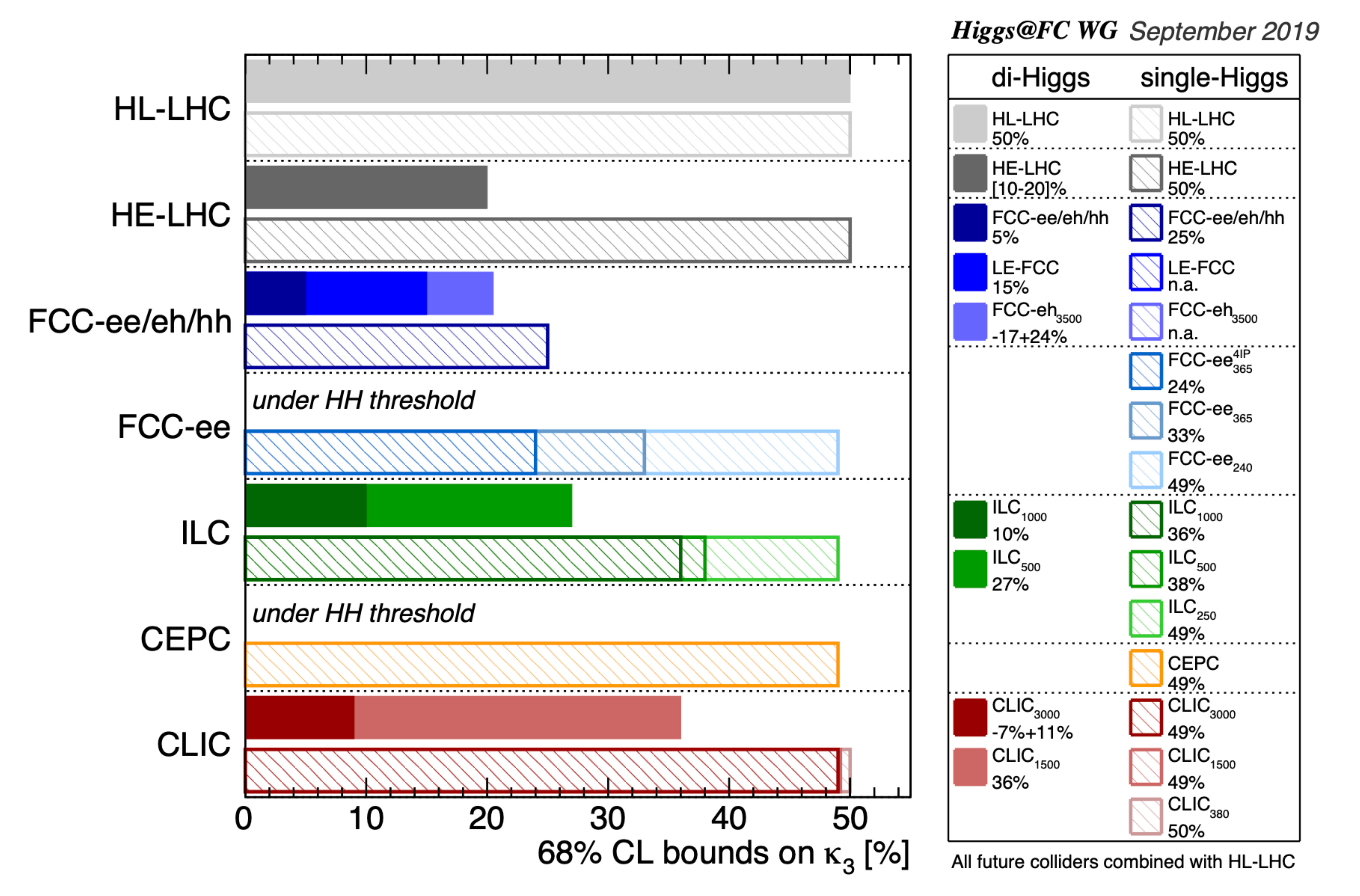}
\caption{Sensitivity at 68\% probability on the Higgs self-coupling parameter $\lahhh$ at the various future colliders. All the numbers reported correspond to a simplified combination of the considered collider with HL-LHC, which is approximated by a 50\% constraint. For each future collider, the result from the single-Higgs from a global fit, and double-Higgs are shown separately; from the Physics Briefing Book~\cite{EuropeanStrategyforParticlePhysicsPreparatoryGroup:2019qin} for ESUPP 2020.} 
\label{fig:hself_hhh}
\end{figure}

Both approaches pose great opportunity and challenges. The state-of-the-art projections on $\lahhh$ following both approaches were summarised in~\cite{EuropeanStrategyforParticlePhysicsPreparatoryGroup:2019qin} as shown in Fig.~\ref{fig:hself_hhh}. (For a broad overview on the Higgs potential at future colliders, see Ref.~\cite{DiMicco:2019ngk}.) To go beyond we propose a list of interesting questions to address by this ECFA study, categorised into three areas.

\subsection*{Theory: beyond the SMEFT}
 \begin{itemize}
  \item How would the existence of extra light scalars impact our perspective? 
  Light additional scalars can contribute at the tree-level (via a BSM triple Higgs coupling) to di-Higgs production and at the loop level have a strong impact on the size of $\lahhh$.
  %In some models that modify the Higgs self-coupling there exist extra light scalars. 
  The direct searches belong to other focus topic, but their existence will invalidate the standard SMEFT approach. How should we then carry out the global interpretation which can provide us a measurement of the Higgs self-coupling?

  \item Provided that the Higgs self-coupling can have ${\cal O}(1)$ deviation, how would this change 
  %dramatically 
  our projections? For instance, if $\lahhh$ is a factor of 2 larger than in the SM, a 500 GeV $e^+e^-$ collider would be able to discover it~\cite{DiMicco:2019ngk}. %\htr{SH: citation?!}
  In general, 
  %we would like 
  it would be important to have projections in both approaches for the cases with non-SM values of $\lahhh$.
\end{itemize}

\subsection*{Single-Higgs: lifting the degeneracies}
The key goal is to isolate out the effects of $\lahhh$ from many others as mentioned above.
\begin{itemize}
 \item Can we lift the degeneracies by employing new observables other than $\sigma_{Zh}$? E.g., using the angular distributions in $Zh$.
 \item Can we take advantage of initial state radiation to realise multiple effective $\sqrt{s^\prime}$ which may help lift the degeneracies? Or how would energy scan just around one nominal $\sqrt{s}$ help?
 \item What if we include other NLO effects as well, e.g. 4-fermion contact interaction from electron and top-quark?
 \item Can we clarify the importance of each input measurement for the $\lahhh$ in the global fit?
 \item Do we expect any update from experimental analyses about single-Higgs observables?
 \item Combination of single-Higgs and double-Higgs sensitivity to $\lahhh$ at $\sqrt{s}\ge$ 500 GeV.
\end{itemize}

The direction of studies related to employing $Zh$ angular observables clearly needs to be in collaboration with one other focus topic on the $Zh$ analysis. It is not very clear what is the best way to incorporate the angular observables in the global analysis for $\lambda_{hhh}$. One may follow the standard template method as that for LHC differential cross section measurement, or compress the angular observables into few effective couplings based on the allowed Lorentz structures in the $hZZ$ couplings, or use the optimal observable approach similar to the strategy for anomalous triple gauge couplings in $e^+e^-\to WW$. 

\subsection*{Di-Higgs production: advancing the analysis technique}

\begin{figure}[h]
	\centering
	\includegraphics[width=\textwidth]{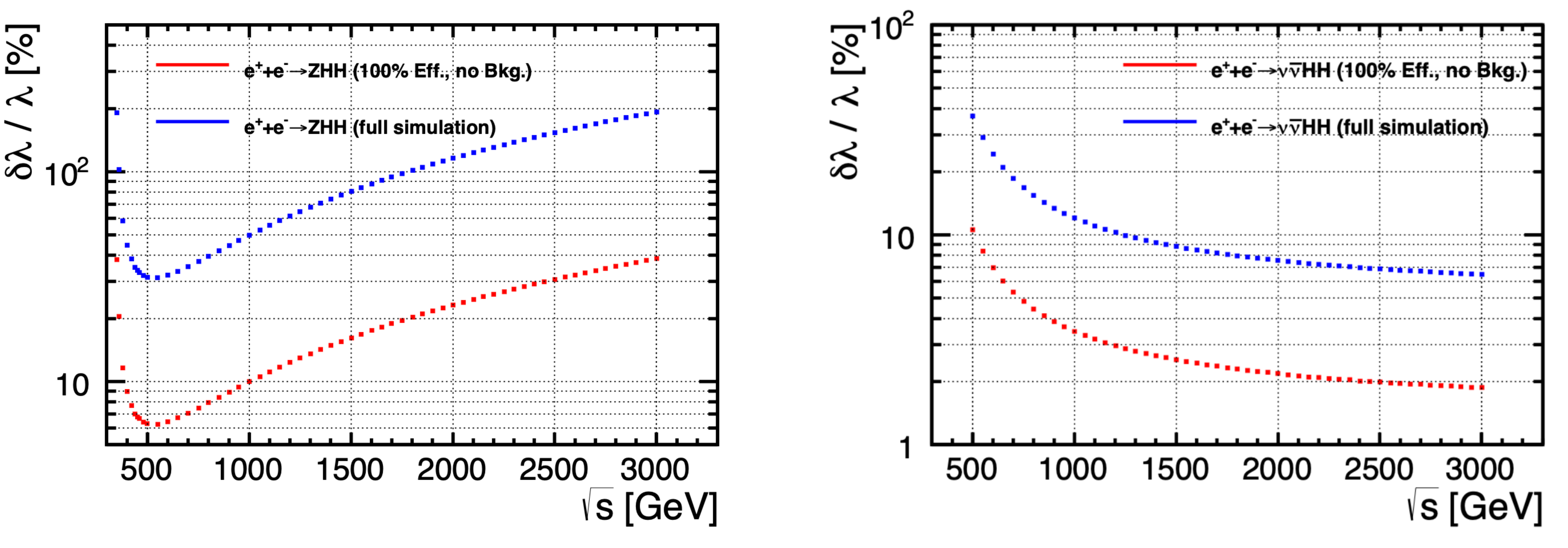}
\caption{Expected precision of $\lambda_{hhh}$ as a function of $\sqrt{s}$ using signal channels $e^+e^-\to Zhh$ (left) and $e^+e^-\to\nu_{e}\bar{\nu_e}hh$ (right), for two scenarios: based on current full simulation analysis (in blue); based on idealised analysis without background and 100\% efficiency (in red).} 
\label{fig:hself_RoomForImprovement}
\end{figure}

The di-Higgs production has a substantially smaller challenge from degeneracies.
The main questions are related to how we can improve experimental analyses either for $\lambda_{hhh}=\laSM$, or for values of $\lahhh \neq \laSM$, or for the case with contributions from new light scalars. 
It has been realised \cite{tian_lcws15} that there is a huge room for potential improvement by comparing the projections based on current full simulation analysis and idealised theoretical expectation (no background and 100\% efficiency) as shown in Fig.~\ref{fig:hself_RoomForImprovement}. The limiting factors in current simulation analysis dominantly come from the algorithms for jet-clustering and flavour tagging. In general more efficient analysis methods which can better suppress background or utilising more signal channels will be helpful, such as employing more sophisticated kinematic fitting~\cite{torndal_lcws23}, matrix element method~\cite{bliewert_ecfa23}, or machine learning~\cite{suehara_ecfa23}, and analysing $Z\to\tau\tau$ signal. Some details about the potential impact to $\lambda_{hhh}$ precision from those improvements can be found in Ref.~\cite{Durig:2016jrs}.

Other than improving the analysis techniques, it would be also interesting to address following questions: 
\begin{itemize}
 \item Would the use of centre-of-mass energies slightly above 500 GeV help the analysis, e.g.\ from more boosted jets?
 \item The event shapes are strongly influenced by the value of $\lahhh$ as well as by tree-level contributions of additional light scalars (where a BSM triple Higgs coupling, here denoted as $\lahhH$ enters the prediction)~\cite{Arco:2021bvf}. 
 \begin{itemize}
\item Can we do some simulation analysis with non-SM value of $\lahhh$?
 \item Can we do some simulations taking into account the effects of $\lahhH$?
\item Can we investigate new event shapes that are more suitable to analyse these effects?
\end{itemize}
\end{itemize}

\subsection*{Target physics observables}
   \begin{itemize}
       {\item Single Higgs observables at 250 + 350/365 GeV:
       \begin{itemize}
           \item for indirect determination of $\kala =\lahhh/\laSM$: EFT approach ($\kala\approx 1$) vs. concrete models ($\kala \not = 1$)
           \item for indirect determination of multi-Higgs interactions involving also extra Higgs bosons. Can these contributions be disentangled from the SM-like Higgs self-coupling? Study the distributions. Is there are an optimal energy scan?
       \end{itemize}
       }
       \item Double-Higgs observables at $\sqrt{s}\leq 500$ GeV: various algorithms that can improve substantially di-Higgs cross section measurements? It is also interesting to study the differential cross section in order to enhance the sensitivity to $\lambda_{hhh}$.
       %\item Di-Higgs production of light BSM Higgses? \htr{SH: interesting of course, but do we want to go there?}\JB{I agree it's interesting but this seems to me outside of the scope of this focus point.}
   \end{itemize}

\subsection*{MC samples needed}
For single-Higgs observables, the MC samples will be needed but taken care of in the $Zh$ focus topic. For di-Higgs observables, we do need MC samples in order to develop new analysis techniques as well as to address impact of $\sqrt{s}$. ILC and CLIC have produced di-Higgs events based on full detector simulation at $\sqrt{s}=500, 1000, 1500, 3000$ GeV. Those samples would become available upon request. There are also publicly available ILC samples produced for Snowmass 2022:~\url{https://pages.uoregon.edu/ctp/ilcsnowmass.html}.
\subsection*{Existing tools / examples}
      \begin{itemize}
          \item Ongoing analysis code for $Zhh$ on GitHub prepared by Julie Torndal:~\url{https://github.com/ILDAnaSoft/ZHH}.
      \end{itemize}

\subsection*{Contact \& Further Information}
\begin{itemize}
\item Gitlab wiki: \url{https://gitlab.in2p3.fr/ecfa-study/ECFA-HiggsTopEW-Factories/-/wikis/FocusTopics/Hself}
\item Sign up for egroup: ECFA-WHF-FT-Hself@cern.ch via \url{http://simba3.web.cern.ch/simba3/SelfSubscription.aspx?groupName=ecfa-whf-ft-hself}
\item and/or email the conveners of ECFA WG1 GLOBal group:\\ \url{mailto:ecfa-whf-wg1-glob-conveners@cern.ch}
\end{itemize}

\clearpage
\section{{\bfseries Wmass} --- Mass and width of the $W$ boson from the pair-production threshold cross section lineshape and from decay kinematics} 
\label{sec:Wmass}
%\title{Focus topics for the ECFA study on Higgs / Top / EW factories}
%\author{A proposal by the WG Coordinators and Subgroup Conveners%\thanks{On leave from another institute somewhere.}}
%\institute{Many...}

%\begin{abstract}
%In order to stimulate new engagement and trigger some concrete studies in areas where further work would be beneficial towards fully understanding the physics potential of a Higgs / Top / Electroweak factory, we propose to define a set of focus topics. The general reasoning and the proposed topics are described in this document.
%\end{abstract}

%\keywords{Higgs / Top / Electroweak Factory, benchmark topics}

%\maketitle
%\tableofcontents
%\setcounter{section}{-1}

%\section{Mass and width of the $W$ boson from the pair-production threshold cross section lineshape and from decay kinematics ($\sqrt{s}=158-380 $\,GeV)} 
\begin{center}
{\itshape Expert Team: Paolo Azzurri, Josh Bendavid, Martin Beneke, Jorge de Blas, Stefan Dittmaier, Ayres Freitas, Adri\'an Irles, Andreas B.\,Meyer, Simon Pl\"atzer, Matthias Schott, Raimund Str\"ohmer, Graham Wilson}
\end{center}

The mass of the $W$ boson ($m_{W}$) is a cardinal parameter of the standard model (SM). 
A precise experimental $m_{W}$  determination is highly desirable to test the SM consistency and as a mean to reveal possible new physics effects~\cite{Haller:2018nnx}.
The level of precision to which  $m_{W}$ could be determined at a future $e^+e^-$ collider with a scan of cross sections at the pair-production  threshold, and from measurements of decay kinematics is greatly superior to current measurements~\cite{Workman:2022ynf}, and expected to by far exceed future measurements at the LHC which are projected to yield uncertainties around 10 MeV~\cite{Dainese:2703572}.

To obtain an estimate of the achievable theory and experimental  systematic uncertainties at a $e^+e^-$ Higgs factory, the potential of analysis, calibration and event reconstruction methods, currently in development, should be assessed. Several complementary methods will be used to measure $m_{W}$. Ideally their systematic uncertainty should match or be better than the statistical precision of the data.
While $W$-pair threshold cross section scans are expected to be less impacted by systematic uncertainties, detector performance will play a crucial role in $m_{W}$ measurements that make use of kinematic information obtained from event reconstruction.
In the following we outline the main methods envisaged.

% Aspects: theory predictions, analysis, calibration and reconstruction strategies. 
\subsubsection*{Pair-production threshold}

The rapid rise of the $W$-pair production cross section near its kinematic threshold can be exploited to obtain a precise and direct determination of the $W$ boson mass.
The threshold method is clean and simple as it just involves classifying and counting events.
A small amount of LEP $e^+e^-$ collision data taken in 1996 at a single energy point near 161~GeV has allowed to determine the the $W$ mass with a precision of 200~MeV~\cite{Barate:1997mn,Abreu:1997sn,Acciarri:1997xc,Ackerstaff:1996nk}.

Collecting over $1~{\rm ab}^{-1}$ of integrated luminosity at threshold energies will enable a determination of the $W$ mass with a statistical uncertainty below 1~MeV~\cite{Abada:2019lih}.
Measuring the threshold cross section at more than one energy point will allow to measure both the $W$ mass and width, with similar precision~\cite{Azzi:2017iih,Azzurri:2021yvl}.
Polarised collisions can enhance or suppress $t$-channel signal production, enabling further control of the background~\cite{Wilson:2016hne,LCCPhysicsWorkingGroup:2019fvj}.

\subsubsection*{\boldmath $W$-pair decay kinematics}
The primary method to measure the $W$ mass and width at LEP2 was with the kinematic reconstruction of semileptonic ($qq\ell\nu$) and fully hadronic ($qqqq$) 
$W$-pair decays~\cite{ALEPH:2006cdc,DELPHI:2008avl,L3:2005fft,OPAL:2002hhr,ALEPH:2013dgf}.
These results were obtained by imposing the constraint that the total four-momentum in the event should be equal to the known initial centre-of-mass energy and zero momentum.

It is foreseen that performing similar measurements with future collider data would yield a statistical precision of a few MeV or less~\cite{LCCPhysicsWorkingGroup:2019fvj,Azzurri:2021yvl},
but the impact of systematic uncertainties is more difficult to predict, in particular those arising from the modelling of non-perturbative QCD effects in the $W$ boson hadronic decays, that stood out in the LEP2 measurements~\cite{ALEPH:2013dgf}.

In the standard (LEP2-style) $W$-mass kinematic reconstruction, non-perturbative QCD uncertainties  arise on two fronts. 
The first is from overall uncertainties on the modelling of core jet properties, in particular on
the jets boost ($\beta_{\rm jet} = p_{\rm jet}/E_{\rm jet}$) that is a key ingredient of the  kinematic fit.  
Uncertainties on $\beta_{\rm jet}$ will affect similarly both the $qq\ell\nu$ and $qqqq$ channels.
The second source is from colour reconnection (CR) effects, that can 
lead to an important impact and uncertainty on $m_{\rm W}$, but only in the $qqqq$ channel. 

The energy spectrum endpoints in leptonic decays,  that are free from QCD uncertainties, can also be used to measure the $W$ mass~\cite{LCCPhysicsWorkingGroup:2019fvj}.
In the case of fully leptonic decays, a pseudomass~\cite{OPAL:2002hhr} can be computed and employed for a $W$ mass measurement.

Finally a direct measurement of the hadronic mass, without kinematic constraints, can be performed~\cite{Anguiano:2020qpk} where the uncertainties will be dominated by the hadronic energy scale.

\subsection*{Theoretical and phenomenological targets and challenges }

\subsubsection*{Precise predictions of total and differential $W$-pair production}
%(Martin Beneke, Stefan Dittmaier, Ayres Freitas, ...  )}
Accounting for the high experimental precision in the $M_W$ determination
at $e^+e^-$ colliders in predictions 
requires a proper inclusion of radiative corrections
to the $e^+e^-\to4f$ processes with intermediate $W$ pairs.
Depending on the $e^+e^-$ CM energy $E_{\mathrm{CM}}$, 
different physical effects and corrections 
dominate, and the target precisions are not the same.
Customised treatments are required for CM energies near the 
W-pair threshold ($|E_{\mathrm{CM}}-2M_W|<n\Gamma_W$, $n\sim2{-}3$),
at intermediate energies of a typical future circular collider
($2M_W+n\Gamma_W < E_{\mathrm{CM}} < 250\mathrm{GeV}$),
or at higher energies.

``Improved-Born approximations''
based on leading universal corrections (initial-state radiation, 
Coulomb singularity, effective couplings)
can be formulated uniformly for the whole energy range, but
are good only to $\sim2\%$ for integrated cross sections
up to intermediate energies (and deliver only $\sim10\%$
accuracy for $E_{\mathrm{CM}}\sim1\mathrm{TeV}$).
For intermediate energies, a pair of on-shell $W$ bosons 
dominates the cross section, so that NLO corrections can be included
in the ``double-pole approximation (DPA)''~\cite{Denner:2000bj,Jadach:2001mp}, which is based on
the leading term of an expansion of amplitudes about the $W$ resonance poles. 
The quality of the DPA was estimated to $0.5{-}1\%$ for integrated
cross sections and later confirmed in a comparison
to NLO predictions for the full $e^+e^-\to4f$ cross sections~\cite{Denner:2005fg} which should be good within
few $0.1\%$ even in the $W$-pair threshold region, where the DPA breaks down
(see Sect.~6.5.5 of \cite{Denner:2019vbn} for more details and 
original references).
At threshold, the currently best calculation is based on complete 
NLO results for $e^+e^-\to4f$~\cite{Denner:2005fg}
and partial higher-order effects 
for the total cross section from an effective field theory (EFT) framework~\cite{Beneke:2007zg,Actis:2008rb}, suggesting a theory uncertainty on $M_W$ of about 3\,MeV for $m_{W}$ extracted from a threshold measurement~\cite{Actis:2008rb}, excluding the uncertainty from QED initial-state radiation.

The quality of differential predictions in dominant phase-space regions 
widely follows the precision estimates for integrated cross sections,
but the precision significantly deteriorates in regions where 
resonant $W$-pair production becomes less dominant, e.g.\ in the
backward direction of $W$ production or at very large CM energies
(see also Sec.~\ref{sec:WWdiff} below and Sec.~8 of Ref.~\cite{Frixione:2022ofv}).

To account for the leap in precision from LEP to a future $e^+e^-$ collider, 
predictions for $W$-pair production have to be improved considerably,
most notably the treatment of initial-state radiation~\cite{Jadach:2019bye,Jadach:2019wol}.
At threshold, NNLO corrections to the ``hard'' $W$-pair production and $W$ decay
processes as well as leading corrections beyond NNLO can be calculated within
the EFT approach, and a precision of 0.5\,MeV in $m_W$ from an
energy scan near threshold seems feasible~\cite{Freitas:2019bre}. 
For intermediate and high energies, a full NNLO calculation for $W$-pair 
production in DPA would be most desirable, a task that should be
achievable in the next decade, anticipating further progress at
the frontier of loop calculations.

\subsubsection*{Estimates and control of QCD effects} %(Simon Plaetzer, Raimund Stroehmer, .. } 

Accurate jet evolution and prediction of radiation patterns are important ingredients to the $W$ mass measurement.

Hadronisation models might impact the jet properties as the boost, and energy scale due to differences in the baryon ratios. For the jet energy calibrations from $Z$ peak events, beam energy, jet mass and jet momentum are affected by baryon and strange ratios.
Hadronisation models can also affect the measured jet energy, resolution and jet mass via the charged/neutral ratio fraction of very low and high $p_{\rm T}$ tracks and the shower shape. $Z$ peak events have a different flavour composition than $W$ bosons. Studies of $b$ and $c$ tags exist but there are also $s$ quarks.

Colour reconnection (CR) in all-hadronic events affects particles with large $\Delta R$ to the jet (parton) direction by pulling them towards or away from the jet, since they change the colour topology underpinning the hadronisation process. These have a large lever arm on the jet direction. At LEP2, the effect was reduced with cuts on $\Delta R$ or cuts on soft particles or weights. Theoretically well-behaved algorithms, e.g.\,Cambridge/Aachen with freeze-out, are needed to address this issue. Experimentally, CR can be accessed using particle flow in the region between jets. Different models with similar measured flow can have significantly different effects on the reconstructed jet direction, and thus the $W$ mass. All ``realistic'' CR models should be considered to estimate the possible impact in lack of further constraints. Constraints on CR can be obtained from change of jet direction (or $W$ mass) by changing parameters of jet algorithms (e.g. jet freeze out). Jet algorithms that do not include soft particles with large $\Delta R$ in the jet associated with the partons from the $W$ boson decay might have larger modelling uncertainties. $Z$ boson peak events with two jets can be used to calibrate the direction resolution.

Newly developed colour reconnection and hadronisation models \cite{Christiansen:2015yqa,Gieseke:2017clv,Gieseke:2018gff,Platzer:2022jny} should be considered to put further constraints on the predictivity for such final states.
   
\subsection*{Detector performance and analysis methods}
%(Graham Wilson, Josh Bendavid, Matthias Schott, Andreas Meyer, ... ) 

Detector and analysis performances are particularly relevant for measurements with the reconstruction of the $W$ boson decay kinematics.
In the case of the reconstruction of fully leptonic or semileptonic $W$ pair decays, the determination of lepton energy scale is a key ingredient. 
%The lepton energy scale is key to the measurement of the $W$ mass, independently of the beam energy, esp.\,when the beam energy is not sufficiently well known event-by-event. 

Previous studies concluded that the evaluation of $J/\psi \to \mu^+\mu^-$ events could provide the most precise constraints of the track momenta, as the $J/\psi$ mass is currently known to the level of 1.9 ppm, corresponding to about 6 keV~\cite{Workman:2022ynf}. However, at $e^+e^-$ colliders the production of $J/\psi$ is expected to be statistics limited. More recent studies reported in the ECFA workshop~\cite{ecfa22_wilson}, based on Refs.~\cite{Madison:2022spc,Rodriguez:2020qhf}, have explored how a combination of mass information from several well measured particles, in particular $K^0_{\rm S} \to \pi^+\pi^-$ and $\Lambda \to p \pi^-$ can be used an absolute calibration of track momenta at a future high-energy $e^+e^-$ collider. With sufficient statistics it is estimated that a centre-of-mass energy uncertainty of $2 \times 10^{-6}$ could be achieved, competitive with the uncertainty in the centre-of-mass energy expected at FCC-ee.

In contrast, the jet energy scale is not expected to be known better than the level of $10^{-4}$, with dominant uncertainties arising from modelling uncertainties such as flavour dependence, hadronisation, colour reconnection, as outlined above.
Nevertheless, measurements of the $W$ boson mass in the hadronic channel, without kinematic constraints, will  provide complementary results with uncorrelated uncertainties, useful for cross check and combination with the measurements in other channels.

Particle reconstruction and flavour identification in jets are important for the event selection and to calibrate and control flavour-dependent effects. In current experiments, novel machine-learning techniques have proven to facilitate unprecedented performance in signal-to-background separation and flavour identification.

If external information of the beam energy is used in the reconstruction of hadronic ($qqqq$) or semileptonic ($qq\ell\nu$) $W$-pair decays, the hadronic jets energy scales can be constrained by the four momentum conservation in the event, removing the importance of the  experimental jet energy determinations. 
Still a precise modelling and control of the hadronic jets boost ($\beta = p/E$) will be essential for precise $m_W$ measurements from $qqqq$ and  $qq\ell\nu$ events. 

In view of necessitating extreme precision on jet internal properties as the boost, it will be highly desirable that the detector capabilities are highly efficient and precise in identifying and measuring all particles inside hadronic jets.  
Excellent particle reconstruction capabilities should extend to low momentum particles away from the jet axis, and in the inter-jet regions, where colour reconnection (CR) effects in the $qqqq$ channel can be determined and constrained {\em in situ}.

An ultimate aim should be to fit simultaneously $WW$, $ZZ$ and $Z\gamma$ leptonic and hadronic  decay modes, in order to extract a determination of the $m_{W}/m_{Z}$ ratio with potentially large cancellations of most systematic uncertainties, from theory, experiment, and beam energy. 
% Detector aspects of centre-of-mass energy scale.

\subsection*{Further work and open questions}
%(Paolo Azzurri, Graham Wilson, Josh Bendavid, ...)

In the following we provide an incomplete list of work packages with possible large impact on the precision estimates. The results can provide important input to future theory developments and detector design.

\subsubsection*{Pair-production threshold}

\begin{itemize}
    
\item Explore in more detail the systematic uncertainties with multi-point ($n\geq 3$) cross section measurements from a $WW$ threshold scan. The reduction and cancellation of point-to-point correlated systematic uncertainties as acceptance, luminosity and background, are of particular interest.

\item Design and implement a modern analysis, using event classifiers based on machine learning and/or a profile likelihood fit with nuisance parameters to describe and constrain systematic uncertainties.

\item Use state-of-the-art generators to evaluate the impact of the various systematic uncertainties and overall performance, for each channel and their combination.

\item Evaluate model-independent approaches in which the theory uncertainties, as outlined above, could be constrained directly from the data.

\item For intermediate and high energies, a full NNLO calculation for $W$-pair  production in DPA would be most desirable, a task that should be
achievable in the next decade, anticipating further progress at
the frontier of loop calculations.

\end{itemize}

\subsubsection*{W-pair decay kinematics}

\begin{itemize}
\item Study a LEP2-style $W$ mass measurement. Estimate the statistical precision of data at different centre-of-mass energies. Study the impact of the systematic uncertainties in detail.

\item Explore simultaneous analysis and fit of diboson events ($WW$, $ZZ$ and $Z\gamma$) to extract $m_W/m_Z$ with potential cancellations of systematic uncertainties both theoretical and experimental. The simultaneous fit of $Z$ peak data for the calibration of lepton and jet energies will also be  necessary.

\item Clarify the usefulness of high-precision cross section measurements in the continuum (see e.g.\,section 4.5 in Ref.~\cite{deBlas:2022ofj}).

\end{itemize}

\subsection*{Monte Carlo samples}
Basic samples for continuum energies are available. Dedicated state-of-the-art samples for threshold scans are needed.

\subsection*{Contact \& Further Information}
\begin{itemize}
\item Gitlab wiki: \url{https://gitlab.in2p3.fr/ecfa-study/ECFA-HiggsTopEW-Factories/-/wikis/FocusTopics/Wmass}
\item Sign up for egroup: ECFA-WHF-FT-WMASS@cern.ch via \url{http://simba3.web.cern.ch/simba3/SelfSubscription.aspx?groupName=ecfa-whf-ft-wmass}
\item and/or email the conveners of ECFA WG1 PRECision group:\\ \url{mailto:ecfa-whf-wg1-prec-conveners@cern.ch}
\end{itemize}

\clearpage
\section{{\bfseries WWdiff} --- Full studies of $WW$ and $e\nu W$} 
\label{sec:WWdiff}
%\section{{\bfseries WWdiff} -- Full studies of $WW$ and $e\nu W$ (JdB)}  
\begin{center}
{\itshape Expert Team: Patrizia Azzi, Timothy Barklow, Jorge de Blas, Ansgar Denner, Alexander Grohsjean, Wolfgang Kilian, Jenny List, Frank Siegert}
\end{center}

\subsection*{Motivation}
Constraints on gauge boson interactions are crucial ingredients to global interpretations, be it in SMEFT or in UV complete models. %, and the search for new physics. 
In particular, in models where the electroweak symmetry is linearly realised in the light fields, new physics contributions to anomalous triple gauge couplings are directly connected to corrections on Higgs couplings, establishing a complementarity between the two sectors of measurements.

\subsection*{Previous experimental studies}
For future $e^+e^-$ colliders, it has been shown from theory-level studies that in principle even the most general set of CP-conserving and CP-violating triple-gauge boson couplings, in total 28 real parameters, can be constrained at a centre-of-mass energy of $500$\,GeV with polarised beams~\cite{Diehl:1997ft, Diehl:2002nj, Diehl:2003qz}. Detector-level simulations have been conducted at energies of $500$\,GeV~\cite{Marchesini:2011aka} and $1$\,TeV~\cite{Rosca:2016hcq}, but considering only the three triple anomalous gauge couplings of the so-called LEP parametrisation. These studies used on LO MC, and were restricted to semi-leptonic $W$-pair events with electrons and muons, i.e.\ they neither included single-$W$ processes, nor events with $\tau$ leptons, nor fully hadronic events. The three couplings and their covariance matrix was then passed on to global interpretations, e.g.\ SMEFT fits. More recently, the formalism of statistical optimal observables for all the CP-even interactions contributing at LO in the dimension-six SMEFT parametrisation 
%the three couplings of the LEP parametrisation 
have been used in global fits~\cite{DeBlas:2019qco,deBlas:2022ofj}, also at lower centre-of-mass energies and including the information from all final states of the $W^+W^-$ decays, but only based on theory-level distributions.

\subsection*{Theory State-of-the-Art}
On the Standard Model side, at LEP2 times, the differential cross section for $W$-pair production including $W$ decays was only known within the double-pole approximation, as implemented in the event generators {\sc YFSWW} \cite{Jadach:2001uu} and {\sc RacoonWW} \cite{Denner:2002cg}. Later, the complete electroweak ${\cal O}(\alpha)$ corrections in the Standard Model were
calculated for the charged-current four-fermion production processes
$e^+e^-\to\nu_\tau\tau^+\mu^-\bar\nu_\mu$,
$u\bar{d}\mu^-\bar\nu_\mu$, and $u\bar{d}s\bar{c}$
\cite{Denner:2005es,Denner:2005fg}. These calculations are available
in an unpublished follow-up code of {\sc RacoonWW} named {\sc
  Racoon4f}. It includes on top of the full ${\cal O}(\alpha)$
corrections also the leading-logarithmic initial-state-radiation
effects beyond ${\cal O}(\alpha)$ in the structure-function approach
(following Ref.~\cite{Beenakker:1996kt} and references therein).  For a more detailed summary on the status of theoretical predictions for $e^+e^- \to W^+W^- \to 4 f$ we refer to Section 8 of Ref.~\cite{Frixione:2022ofv} and the discussion in Sec.~\ref{sec:Wmass}.

Predictions in SM extensions like the dimension-six SMEFT are doable thanks to UFO models like, e.g. {\sc SMEFTsim}~\cite{Brivio:2017btx,Brivio:2020onw} at LO, which can be used in {\sc MadGraph}~\cite{Alwall:2014hca}. Automated calculation of NLO corrections in QCD is doable via the UFO model {\sc SMEFT@NLO} \cite{Degrande:2020evl}. The automated calculation of NLO electroweak corrections will be completed and available in codes like {\sc MadGraph} and {\sc Whizard}~\cite{Kilian:2007gr}. Such electroweak corrections are expected to be large at high energies accessible to some of the $e^+e^-$ collider projects under consideration.

\subsection*{Goals of this Focus Topic}
Therefore the main objective of this focus topic is to understand the full potential of $e^+e^-$ colliders with respect to gauge boson interactions, using the full differential information from $W$-pair and single-$W$ events to extract CP-even and CP-odd couplings, based on detailed detector simulation with assessments of systematic uncertainties, at all centre-of-mass energies. 

It is also important to establish the complementarity of these studies with similar studies of anomalous triple gauge couplings (aTGC)  at the HL-LHC and to clarify gain in precision that is expected at future $e^+ e^-$ colliders.

\subsection*{Proposed Study Targets}
\begin{itemize}
    \item Detector-level projections for differential cross-section measurements with respect to production and decay angles, at all centre-of-mass energies with and without beam polarisation.
    \item Detector-level projections for optimal observables for CP-even and CP-odd anomalous gauge couplings, again at all centre-of-mass energies with and without beam polarisation.
    \item Reduction of systematic uncertainties by inclusion of nuisance parameters in the combined interpretation of various data sets.
    \item Estimation of residual systematic uncertainties and their incorporation in the optimal observable formalism.
    \item Inclusion in the studies of single-$W$ production, to establish its role of in constraining electroweak interactions, when combined with di-boson production.
\end{itemize}

\subsection*{MC samples needed}
Some basic samples are available as listed in the Motivation Section, dedicated samples with higher statistics and in particular with anomalous couplings might be needed.

\subsection*{Existing tools / examples}
\begin{itemize}
    \item ILD $qq l\nu$ analysis \url{https://github.com/ILDAnaSoft/ILDbench_WWqqlnu}
\end{itemize}

\subsection*{Contact \& Further Information}

\begin{itemize}
\item Gitlab wiki: \url{https://gitlab.in2p3.fr/ecfa-study/ECFA-HiggsTopEW-Factories/-/wikis/FocusTopics/WWdiff }
\item Sign up for egroup: ECFA-WHF-FT-WWdiff @cern.ch via \url{http://simba3.web.cern.ch/simba3/SelfSubscription.aspx?groupName=ecfa-whf-ft-wwdiff }
\item and/or email the conveners of ECFA WG1 GLOBal group:\\ \url{mailto:ecfa-whf-wg1-glob-conveners@cern.ch}
\end{itemize}

\clearpage
\section{{\bfseries TTthres}  --- Top threshold: Detector-level simulation studies of $e^+e^- \to t \bar{t}$ and thres\-hold scan optimisation} 
\label{sec:TTthres}
% mainfile: ../benchmarks.tex

%\section{{\bfseries {\bfseries TTthres} } -- Top threshold: Detector-level simulation study of $e^+e^- \to t \bar{t}$ at a typical threshold-scan energy ($\sqrt{s}=350, 365$ \,GeV) and threshold scan optimisation (JL) } 
\begin{center}
{\itshape Expert Team: Patrizia Azzi, Martin Beneke, Jorge de Blas, Matteo Defranchis, Gauthier Durieux, Roberto Franceschini, Andr\'e Hoang, Adri\'an Irles, Abideh Jafari, Victor Miralles, Zohreh Najafabadi, Laura Pintucci, Andrej Saibel, Reinhard Schwienhorst, Frank Simon, Marcel Vos, Aleksander Filip Zarnecki}
\end{center}

\noindent
A scan of the $e^+e^-$ centre-of-mass energy through the top quark pair production is part of the programme of all proposed Higgs/top/EW factories. The shape of the top quark pair production rate versus centre-of-mass energy is sensitive to the top quark mass and width, while the absolute cross section around the threshold is sensitive to the strong coupling and the top quark Yukawa coupling. Measurements at centre-of-mass energies above the $t\bar{t}$ threshold can yield precision measurements of the top quark electroweak couplings and have exquisite sensitivity to new physics scenarios where these are altered.

The earliest proposals for a threshold scan\cite{Gusken:1985nf, Strassler:1990nw, Guth:1991ab, Jezabek:1992np} as the "ultimate" top quark mass measurement go back to the 1980s and 1990s. Since then, considerable progress has been made towards precise predictions of the total cross section in the threshold region, which is now available at NNNLO \cite{Beneke:2015kwa}, 
and (partial) NNLL accuracy \cite{Hoang:2013uda}. A code is available that provides NNNLO prediction of the cross section\cite{Beneke:2016kkb} (with the option to include QED initial-state radiation), which has been used extensively for experimental studies. Work towards a generator that includes effect of the Coulomb pseudo-bound-state cross-section enhancement\cite{Bach:2017ggt} is ongoing. 

After early experimental studies by Martinez \& Miquel\cite{Martinez:2002st}, the current strategy follows work by Seidel et al.~\cite{Seidel:2013sqa}. A 10-point threshold scan is envisaged, with points spaced by 1~GeV. The total integrated luminosity is 100\:fb$^{-1}$~\cite{CLICdp:2018esa}.

The expert team aims to provide a firm basis for the projected precision of the top quark mass and width measurements, including a realistic estimate of systematic uncertainties from theory and experiment. This includes, importantly, a full-simulation study of the top quark cross-section measurement at the threshold. Detector-level studies of top physics exist mainly at centre-of-mass energies of $380 - 500$\,GeV and above, focusing on precision determinations of the electroweak couplings of the quarks. Threshold scans have been studied so far based on toy measurements of the total cross-section near threshold. These considered simultaneous extractions of top mass, width, Yukawa coupling and strong coupling $\alpha_{\rm s}$, but included neither differential information nor polarisation.

Discussion among theorists and experimentalists are needed to establish exactly which corrections and cuts to the data are expected to be applied by the experimental collaboration, and which effects can instead be included in the theoretical prediction.

Finally, the expert team hopes to provide a perspective for the top quark mass measurement by comparing the expected precision to the HL-LHC prospects, and by embedding the top quark prospects in the global electroweak precision fit, together with the projections for other measurements, such as that of the $W$-boson mass, the strong coupling, etc. 

Beyond the top quark mass, studies of the top quark width measurement from a threshold scan should be compared to projections of HL-LHC and to the predictions of new physics scenarios that involve non-standard "invisible" decays of the top quark. New ideas are required to turn the extraction of the strong coupling and the top quark Yukawa coupling from the threshold scan into competitive results.

The impact of top quark coupling measurements in operation of the $e^+e^-$ collider above threshold has been studied in detail~\cite{Durieux:2018tev,Bernreuther:2017cyi,Amjad:2015mma}. The $e^+e^-$ prospects are compared to current LHC results and expected HL-LHC bounds on operator coefficients involving top quarks in a global fit of the top sector of the SMEFT~\cite{Schwienhorst:2022yqu,deBlas:2022ofj,Durieux:2019rbz}. These studies are being extended to include a complete NLO parametrisation of the dependence of top quark production processes on all operator coefficients for hadron colliders. Future work may include extension to the imaginary parts of the relevant Wilson coefficients and --- eventually --- a complete merger of the SMEFT fits of the top sector with the Higgs/EW fits.  

\subsection*{Theoretical and phenomenological targets}
\begin{itemize}
    \item Complete and harmonised assessment of systematic uncertainties on SM parameters extracted from the threshold scan.  
    \item Degeneracies in a EFT analysis including only ``one'' energy point.  How to disentangle effects combining with other (non-top-quark) measurements.  Indirect constraints on top Yukawa.
\end{itemize}

\subsection*{MC samples needed}
Basic samples available as listed in the Motivation Section, dedicated samples for threshold scan are needed.

\subsection*{Existing tools / examples}
\begin{itemize}
    \item ILD $t\bar{t}$ analysis \url{https://github.com/ILDAnaSoft/ILDbench_QQbar}
    % what's that? Reference? Link?
    %\item QQthreshold 
\end{itemize}

\subsection*{Contact \& Further Information}

\begin{itemize}
\item Gitlab wiki: \url{https://gitlab.in2p3.fr/ecfa-study/ECFA-HiggsTopEW-Factories/-/wikis/FocusTopics/TTthresh }
\item Sign up for egroup: ECFA-WHF-FT-TTthres@cern.ch via \url{http://simba3.web.cern.ch/simba3/SelfSubscription.aspx?groupName=ecfa-whf-ft-ttthres }
\item and/or email the conveners of ECFA WG1 GLOBal group:\\ \url{mailto:ecfa-whf-wg1-glob-conveners@cern.ch}
\end{itemize}

\clearpage    

\section{{\bfseries LUMI} --- Precision luminosity measurement} 
\label{sec:LUMI}
%\section{{\bfseries LUMI} Precision of the luminosity measurement from low-angle Bhabha scattering (JL)} 
\begin{center}
%{\itshape Expert Team: Ivanka Bozovic, Mogens Dam, Fulvio Piccinini, Wies\l aw P\l aczek, Andr\'e Sailer, Maciej Skrzypek, Graham Wilson; Paolo Azzuri, Ayres Freitas, Adri\'an Irles, Andreas B.\,Meyer}
{\itshape Expert Team: Paolo Azzuri, Ivanka Bozovic, Mogens Dam, Ayres Freitas, Adri\'an Irles, Andreas B.\,Meyer, Fulvio Piccinini, Wies\l aw P\l aczek, Andr\'e Sailer, Maciej Skrzypek, Graham Wilson }
\end{center}

Precision measurements of the luminosity are important for all cross-section and line-shape measurements, in particular the $Z$ peak cross section, $\sigma_Z^0$; the total $Z$ width from the line-shape of $e^+e^- \to f\bar{f}$; the $W$ boson mass and width from the line-shape of the cross-section for $e^+e^- \to W^+W^-$ near threshold; the total cross section for $e^+e^- \to HZ$ (used for extracting the effective $HZZ$ coupling and the total Higgs boson width). At LEP, an absolute calibration of the luminosity with a relative experimental uncertainty of $3.4 \times 10^{-4}$ has been achieved \cite{OPAL:1999clt}, using small-angle Bhabha scattering. For future $e^+e^-$ colliders, the luminosity uncertainty will likely be the limiting factor for several of the measurements listed above, in particular on the $Z$ pole, so it is crucial to reduce the uncertainty compared to LEP.

A realistic target for the overall luminosity calibration at the $Z$ pole is $10^{-4}$ or better, whereas for the point-to-point luminosity control, i.e.\ the relative uncertainty between two nearby centre-of-mass energies or two beam polarisation settings, one would like to reach $\mathcal{O}(10^{-5})$ precision (see e.g.\ Ref.~\cite{FCC:2018evy}). 
At intermediate energies, $W^+W^-$ and two-fermion production are the 
highest cross-section processes of interest leading 
to anticipated data-sets of $\mathcal{O}(10^{7} - 10^{8})$ such events thus 
motivating a similar $\mathcal{O} (10^{-4})$ experimental target for luminosity precision.
In particular to obtain a precision below 1~MeV on the $W$ boson mass from the $W$-pair production lineshape at threshold, a control of the luminosity uncertainty at the level of few $10^{-4}$  will be needed~\cite{Azzi:2017iih,Azzurri:2021yvl}.
At higher energies, $\sqrt{s} \geq 400$~GeV, $\mathcal{O}(10^{-3})$ precision for the overall luminosity calibration may suffice for the physics goals \cite{BozovicJelisavcic:2013lni}.

\medskip\noindent
The physics processes used for luminosity calibration need to meet certain requirements:
\begin{itemize}
    \item ideally large rate, so as not to be statistics limited also in small time intervals, and low backgrounds;
    \item good control of the experimental systematic uncertainties (particle ID, acceptance);
    \item reliable, high-precision theory predictions and MC tools, with negligible room for possible new physics contributions.
\end{itemize}

In the following, we summarise the state-of-the-art concerning two processes fitting this bill, namely small-angle Bhabha scattering (SABS) and di-photon production, before summarising the most important open questions.

\subsection*{Small-angle Bhabha scattering (SABS)}
The cross section for Bhabha scattering is very strongly peaked in the forward region, so that the best rate measurement can be performed with a special detector (LumiCal) at $< 100$\,mrad. Experimental challenges at future $e^+e^-$ colliders are numerous, and systematic uncertainties drive the precision of the integrated luminosity measurement. Precision is particularly critical for the $Z$ pole line-shape measurement. Systematic effects comprise detector-related uncertainties, beam-related uncertainties and uncertainties originating from physics and machine-related interactions. They should be preferably quantified in a full detector simulation including backgrounds in the very forward region. This calls for novel and revised studies, in particular at linear colliders, in line with the evolving design of the MDI region.

%\medskip\noindent
\subsubsection*{Experimental challenges:}
\begin{itemize}
    \item Detector (luminometer) aperture, position and alignment; some of the requirements for mechanical precision are more stringent than others (i.e. inner radius), posing technological challenges for detector realisation; the uncertainties at the ILC are discussed in Ref.~\cite{stahl2005luminosity} for the 500\,GeV run and should be revisited for the latest proposed detector design~\cite{theildconceptgroup2010international} and ILC operating scenarios~\cite{Barklow:2015tja}. At FCC-ee, the design of the MDI region requires the luminosity monitor to be placed closer to the IP compared to LEP or ILC, which puts even higher requirements on the geometrical precision for the same angular acceptance uncertainty \cite{Dam:2021sdj}. The following table summarises the requirements for how well the position and dimensions of the luminosity detector need to be controlled. The cited uncertainties include changes between offline metrology and operations, which was a dominant uncertainty for the inner radius acceptance at OPAL~\cite{OPAL:1999clt}.

    \smallskip
    \begin{tabular}{|l|c|c|c|}
    \hline
         & LEP \cite{OPAL:1999clt} & FCC-ee ($Z$ pole) & ILC  \cite{stahl2005luminosity}, \cite{theildconceptgroup2010international} \\ 
         & & &  ($\sqrt{s} > 250$~GeV)  \\
         \hline
         LumiCal distance from IP [m] & 2.5 & 1.1 & 2.48 \\
         \hline
         Precision target & $3.4 \times 10^{-4}$ & $10^{-4}$  & $10^{-3}$ \\
         \hline
         Tolerance for &&& \\
         ~~~inner radius [$\mu$m] & 4.4 & $\mathcal{O}(1)$ & 4 \\
         ~~~outer radius [$\mu$m] & ? & $\lesssim 3$ & ? \\
         ~~~distance between two LumiCals [$\mu$m] & $\mathcal{O}(100)$ & $<100$ & 200 \\
         \hline
    \end{tabular}
    \medskip

   % The cited uncertainties include changes between offline metrology and operations, which was a dominant uncertainty for the inner radius acceptance at OPAL \cite{OPAL:1999clt}.

    \item Beam properties and delivery to the interaction point (IP); beam-energy asymmetry, energy calibration, IP displacements due to finite transverse beam size and beam synchronisation, beam-spread effects. This is yet to be quantified at linear colliders, while at circular colliders it has been discussed in Ref.~\cite{Smiljanic:2020wvt}. Energy calibration is important because the selection of Bhabha events over background (e.g. from two-photon processes) requires accurate calibration of the LumiCal energy scale. In addition, the Bhabha scattering rate depends on the beam energy, and thus the beam energy uncertainty propagates to the luminosity uncertainty (a potential limiting factor for linear colliders, discussed in Ref.~\cite{Lukic:2013fw}).

    \item Machine and physics background; the issue of machine background from the incoherent photon conversion to $e^+e^-$ pairs is of importance at linear colliders where beamstrahlung is a relevant source of photons influencing the luminometer occupancy, in particular at higher centre-of-mass energies. Two-photon (Landau--Lifshitz) process as a source of physics background should also be considered. At linear colliders this is discussed in Ref.~\cite{Abramowicz:2010bg}.
    \item Beam-beam interactions on Bhabha; these comprise beamstrahlung modifying the differential rate of Bhabha scattering and electromagnetic deflection, the latter being pronounced at lower centre-of-mass energies ($Z$ pole). The effects have been studied at linear colliders~\cite{BozovicJelisavcic:2013lni,Lukic:2013fw} and at FCC-ee~\cite{Voutsinas:2019hwu}. Focusing of final state particles: $\mathcal{O}(10^{-3})$ correction due to scattered $e^\pm$ propagating through beam bunches~\cite{Voutsinas:2019hwu}; at Higgs factories this becomes more complicated due to the finite beam-crossing angle, but at the same time it opens the opportunity to measure the focusing effect through the acollinearity distribution of Bhabha events.
    
\end{itemize}

%\medskip\noindent
\subsubsection*{Theoretical challenges \cite{Jadach:2018jjo,Jadach:2021ayv}:}
\begin{figure}
\begin{subfigure}{0.395\textwidth}
\begin{center}
\includegraphics[width=0.50\hsize]{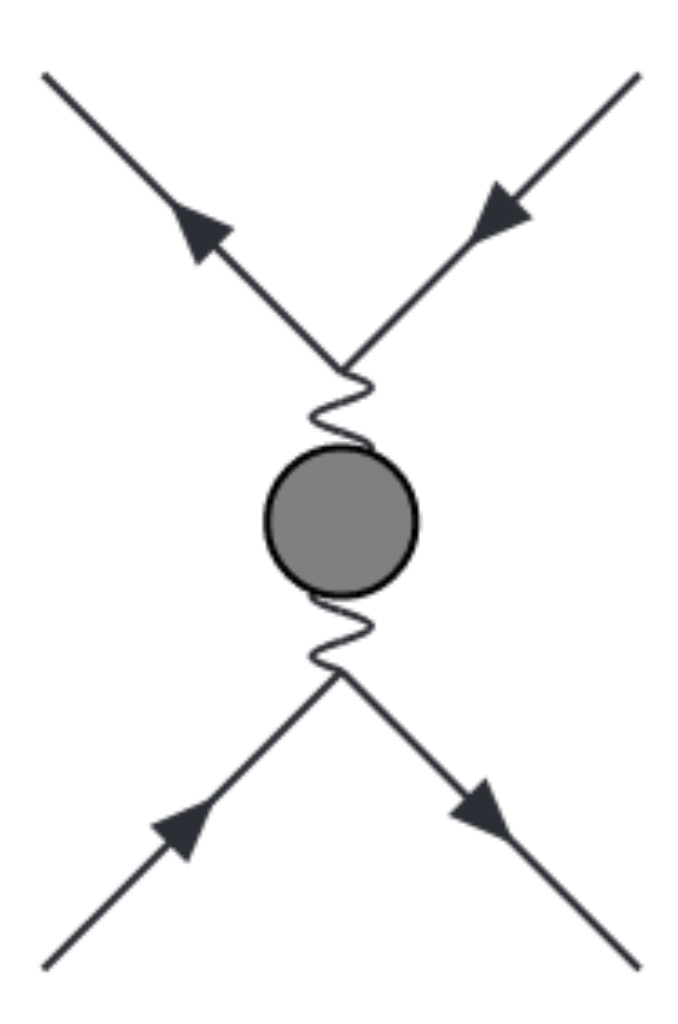}
\end{center}
\caption{\label{fig:lumi:vacpol}}
\end{subfigure}
\begin{subfigure}{0.595\textwidth}
\begin{center}
\includegraphics[width=0.50\hsize]{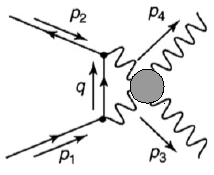}
\end{center}
\caption{\label{fig:lumi:lbl}}
\end{subfigure}
\caption{(a) Vacuum polarisation contribution to Bhabha scattering (from Ref.~\cite{Janot:2019oyi}). (b) Light-by-light correction to di-photon production.}
\label{fig:lumi}
\end{figure}

\begin{itemize}
\item Bhabha scattering is mostly a QED process, i.e.\ higher order corrections can be reliably calculated. Implementation of these corrections in MC tools is complicated but not a fundamental obstacle.
\item Production of additional fermions has a significant impact on the simulated LumiCal Bhabha rates. The technology for computing 4-fermion processes at NLO (see \mbox{e.g.}~\cite{Denner:2005fg}) and 6-fermion processes at LO exists, but these still need to merged in a coherent MC simulation. Inclusion of these contributions should reduce the uncertainty from fermion pair production below $10^{-4}$.
\item Hadronic vacuum polarisation in t-channel photon exchange (Fig.~\ref{fig:lumi:vacpol}). This contribution needs to be extracted from data for $e^+e^- \to \text{hadrons}$ or from lattice QCD. With future data (Belle~II, BES III, CMD-3, SND) it is expected that the uncertainty can be reduced below the $10^{-4}$ level~\cite{Jadach:2018jjo}, but it may be a limiting factor in the achievable precision.
\item NLO electroweak corrections are missing in existing Bhabha MC tools, but they are straightforward to implement.
\item Corrections from linear photon emission and EW higher orders are enhanced at higher energies, thus increasing the theory uncertainty for the luminosity determination there. However, they stay safely below the $10^{-3}$ level for $\sqrt{s}$ up to 1~TeV~\cite{Jadach:2021ayv}.
\end{itemize}

%\medskip\noindent
\subsubsection*{Available MC tools:}
\begin{itemize}
    \item {\sc BHLUMI 4.04} \cite{Jadach:1996is}: NLO QED corrections; $\alpha^2L^2$ and $\alpha^2L$ corrections ($L=
    \ln(|t|/m_e^2)$); higher-order QED corrections through the exclusive YFS exponentiation; no EW corrections; partial implementation of s-channel $\gamma/Z$ exchange; no fermion-pair production.
    \item {\sc BHWIDE 1.00} \cite{Jadach:1995nk}: dedicated to large-angle Bhabha scattering, not optimised for SABS; NLO QED and EW corrections, including $Z$-exchange contributions;  higher-order QED corrections through the exclusive YFS exponentiation; no fermion-pair production.
    \item {\sc BabaYaga} \cite{Balossini:2006wc}: NLO QED corrections, including corrections to $Z$-exchange contributions; higher-order leading-log resummation through parton shower; no EW corrections; no fermion-pair production.
    \item {\sc MCGPJ} \cite{Arbuzov:2005pt}: NLO QED corrections; higher-order leading-log resummation through QED structure functions.
    \item {\sc McMule} \cite{Banerjee:2020rww,Banerjee:2021mty}: NNLO fixed-order corrections.
\end{itemize}

\subsection*{Di-photon production}
Di-photon production ($e^+e^- \to \gamma\gamma$) is also a QED-dominated process that avoids some of the challenges of SABS, in particular the severe metrology requirements and the significant impact of the hadronic vacuum polarisation.

%\medskip\noindent
\subsubsection*{Experimental challenges:}
\begin{itemize}
\item Statistical precision: Di-photon production is also peaked at small angles, albeit to a lesser extent than Bhabha scattering. Nevertheless, the event selection limits the analysis to the central tracker region ($|\cos\theta| \lesssim 0.9$), where track rejection can be applied. This, in turn, limits the statistical precision to about $5 \times 10^{-5}$ with 10~ab$^{-1}$ at the $Z$ pole and about $4 \times 10^{-4}$ with 5\,ab$^{-1}$ at $\sqrt{s} = 250$~GeV.
\item Background: The cross-section from Bhabha scattering is 100 times larger. For an overall luminosity precision of $10^{-4}$ from di-photon production, Bhabha needs to be reduced by a factor $10^{6}$, i.e.\ a factor $10^{3}$ per track (this should be doable in central tracking region). Backgrounds from neutral pion production are expected to be very small.
\item Acceptance/metrology: For $10^{-4}$ precision, the angular acceptance needs to be determined to 50~$\mu$rad which is much looser than the acceptance precision required for SABS. However, here this is needed for the central detector, which consists of several components, with potential cracks, etc. 
\end{itemize}

%\medskip\noindent
\subsubsection*{Theoretical challenges:}
\begin{itemize}
\item The photon vacuum polarisation only appears one order higher than in SABS, i.e.\,at NNLO and thus its uncertainty is negligible~\cite{CarloniCalame:2019dom}. However, there are also hadronic light-by-light (lbl) scattering corrections (Fig.~\ref{fig:lumi:lbl}), which also enter at NNLO, but which have a much larger relative uncertainty (about 20\% for the muon magnetic moment). Their impact on $\gamma\gamma$ production has not been studied. The lbl contribution can be estimated from hadronic models or lattice QCD.
\item Due to the large-angle requirement ($|\cos\theta| \lesssim 0.9$), EW corrections are relatively more important than for SABS~\cite{CarloniCalame:2019dom}. By including NLO and leading NNLO EW corrections, the uncertainty from higher-order EW corrections can likely be rendered sub-dominant, but this requires further investigation.
\end{itemize}

\subsubsection*{Available MC tools:}
\begin{itemize}
    \item {\sc BabaYaga} \cite{Balossini:2008xr,CarloniCalame:2019dom}: NLO QED+EW corrections; higher-order leading-log resummation through parton shower; no fermion-pair production.
    \item {\sc BKQED} \cite{Berends:1980px}: NLO QED corrections; no parton shower or other effects.
    \item {\sc McMule} \cite{Banerjee:2020rww}: NNLO fixed-order corrections.
\end{itemize}

\subsection*{Open Questions}
In the following we point out a few concrete questions which should be addressed with high priority:
\begin{enumerate}
    \item Di-photon production may be a promising candidate for a robust high-precision determination of the total integrated luminosity. A detailed study of the luminosity calibration using this process is still lacking and would be very important.
    \item SABS is preferred for the point-to-point luminosity control, due its higher yields. Some leading systematic uncertainties should drop out in the point-to-point comparison. More studies on this last point are needed, in particular for the understanding of correlations between luminosity measurements at different centre-of-mass energies.
    \item Detailed designs for LumiCal detectors are needed for different collider setups and different detector concepts.
    \item Radiative production of additional fermion pairs is currently not implemented in typical MC programs for SABS and $\gamma\gamma$ production. What is their impact in the experimental analysis of the luminosity measurements?
    \item What is (quantitatively) the impact of beamstrahlung on the overall luminosity determination? Will the beamstrahlung spectrum need to be obtained from simulation, or can be determined from in-situ measurements?
    \item Are there other processes besides $e^+e^-\to e^+e^-$ and $e^+e^- \to \gamma\gamma$ that could be useful for luminosity measurements?
\end{enumerate}

\subsection*{Contact \& Further Information}
\begin{itemize}
\item Gitlab wiki: \url{https://gitlab.in2p3.fr/ecfa-study/ECFA-HiggsTopEW-Factories/-/wikis/FocusTopics/LUMI}
\item Sign up for egroup: ECFA-WHF-FT-LUMI@cern.ch via \url{http://simba3.web.cern.ch/simba3/SelfSubscription.aspx?groupName=ecfa-whf-ft-lumi}
\item and/or email the conveners of ECFA WG1 PRECision group:\\ \url{mailto:ecfa-whf-wg1-prec-conveners@cern.ch}
\end{itemize}

\clearpage
\section{{\bfseries EXscalar} --- New exotic scalars}
\label{sec:EXscalar}
%\section{{\bfseries EXscalar} New exotic scalars (JL)}

\begin{center}
{\itshape Expert Team: Mikael Berggren, Sven Heinemeyer, Abdollah Mohammadi, Tania Robens, Nikolaos Rompotis, Aleksander Filip Zarnecki}
\end{center}

The physics program of the Higgs factory will focus on measurements of the 125\,GeV Higgs boson, with the Higgs-strahlung process being the dominant production channel at 240/250\,GeV \cite{Dawson:2022zbb}. % Snowmass Higgs report 
%
% JL too defensive?
% However, current searches still leave room for additional scalar states. 
However, additional scalar states, especially light ones with masses of the order of, or even below the mass of the Higgs boson observed at the LHC, are well motivated theoretically~\cite{Heinemeyer:2021msz,Biekotter:2022jyr,Robens:2022zgk} and by far not excluded experimentally. Typically, the couplings of such a new state are modified with respect to the SM prediction (for a "SM Higgs boson" at that mass) %SH: not sure whether you want to say this
and highly depend on the specific theory. Due to sum rules stemming from unitarity, for % most of the 
many of the possible
additional CP-even neutral states the couplings to electroweak gauge bosons are suppressed with respect to the SM couplings, see e.g. discussion in Ref.~\cite{Gunion:1990kf}.

Higgs factories can be sensitive to exotic scalar production even for very light scalars, thanks to clean environment, precision and hermeticity of the detectors.
Different production and decay channels, including invisible decays, can be considered. Comprehensive summaries for searches for such additional scalars in various production and decay modes at LEP are documented in Refs.~\cite{OPAL:2002ifx,ALEPH:2006tnd}. Observing and measuring the properties of additional scalars also allows to test the nature of electroweak phase transition, see e.g. discussions in Refs.~\cite{Carena:2022yvx,Papaefstathiou:2022oyi}.
In the following, we propose two production modes for extra scalars which deserve particular attention.

\subsection*{Scalar-strahlung}
Higgs factories are best suited to search for light exotic scalars, here denoted as $\phi$, in the process similar to the Higgs-strahlung:
\[ e^+ e^- \to Z  \phi~. \]
As for the SM Higgs boson, the production of new scalars can be tagged, independent of their decay, based on the recoil mass technique \cite{Yan:2016xyx}. 
Similar analysis methods can be used, looking for corresponding light scalar decay channels (e.g. $b\bar{b}$, $W^{(*)+}W^{(*)-}$, $\tau^+\tau^-$ or invisible decays), but relaxing the constraint imposed by the SM Higgs boson mass. 
Non-standard decays channels of the new scalar (e.g. decays to long-lived particles) can also be looked for, similarly as they should be addressed for the 125\,GeV Higgs. 
For maximum sensitivity, the feasibility of including hadronic $Z$ decays should be explored. 

\subsection*{Higgs decays to Extra Scalars}
In addition to the associated production of the new scalar with the $Z$ boson, other production channels can also be considered, including production processes involving decays of heavier particles (SM gauge bosons, Higgs boson, top quark or new exotic particles).
As as second benchmark scenario for the EXscalar focus topic, 
light scalar pair-production in 125\,GeV Higgs boson decays is proposed: 
\[ e^+ e^- \to Z  H \to Z  \phi \phi \; . \]
Here again, different decay channels of the new scalar state should be considered, both SM-like decays (e.g. $b\bar{b}$, $\tau^+\tau^-$) and exotic ones (e.g. invisible decays) should be considered.
While new scalar states could in general be long-lived, only scenarios with prompt decays are included in this focus topic (while a dedicated topic focuses on LLPs). 

\vspace{-0.23cm}

\subsection*{Target Physics Observables}

The first target in either of the above cases is to quantify the parameter space accessible for discovery (or exclusion) in terms of mass of the scalar and its mixing with the 125\,GeV Higgs boson. However, one should also address the prospects for scrutinising the new scalar state(s) once discovered. How accurately can its properties can be measured? To which extend can these measurements then be used to discriminate theoretical models proposed to describe the excess? For the latter, the properties of the extra scalar need to be combined with the 125\,GeV Higgs boson coupling measurements and other precision measurements at Higgs factory, as considered in the past e.g.\ in Ref.~\cite{Fujii:2017vwa}.

\subsection*{Target detector performance aspects and reconstruction methods to be developed}

Many reconstruction algorithms for the studies proposed here exist in {\sc Key4hep}. In the areas below, performance improvements are expected from developments of more sophisticated approaches: %, including machine-learning:
\begin{enumerate}
    \item Recoil mass reconstruction is the primary method for identification of the SM-like Higgs boson production events and can also be used for new scalar searches. In fact, precision of the recoil mass reconstruction is expected to be the leading factor determining the search sensitivity in the scalar-strahlung scenario, as well as for the exotic Higgs decays.
    \item Resolution in the reconstruction of the invariant mass from the hadronic decays of the produced heavy objects ($Z$ boson, Higgs boson, exotic scalar) is also very important.
    \item Invariant-mass reconstruction can also be improved, using the appropriate corrections for semi-leptonic decays of  heavy flavours, for scalar decays to $b\bar{b}$ and $c\bar{c}$ \cite{Radkhorrami:2021fhe}.
    \item Reconstruction of the invariant mass for scalar decays to $\tau^+\tau^-$ is the key issue for scenarios where these decays dominate (see e.g.\ Ref.~\cite{Biekotter:2022jyr}). This has already been considered in the past for the SM Higgs boson \cite{Tian:2016qlk}, but should probably be developed further. 
    \item Finally, efficient tagging of ISR photons is also important for proper reconstruction of event kinematics and background suppression.
\end{enumerate}

\vspace{-0.23cm}

\subsection*{MC samples needed}

Samples of the study of new scalar production in the scalar-strahlung process, assuming SM-like decays of the new particle ($b\bar{b}$, $W^+W^-$, $\tau^+\tau^-$, $\gamma \gamma$, \dots) can be easily generated using the Standard Model simulation with modified Higgs boson mass, and usually the accordingly modified width.

For the simulation of Higgs-strahlung events with a 125\,GeV Higgs boson decaying to a pair of new scalars an arbitrary model can be selected, assuming it supports the considered decay channels of the new scalar. 
%
%%%%%%%%%%%%%%%%%%%%%%%%%%%%%%%%%%%%%%%%%%%%%%%%%%%%%%%%%%%%%%%%%%%
Angular distributions of the decay products do not depend on the model details and the final sensitivity limits will be given in terms of the cross section times branching ratio.% anyway.

\subsection*{Existing tools / examples}

Prospects for observing new scalar production in the scalar-strahlung process were previously studied for ILC at 250\,GeV on theory level~\cite{Drechsel:2018mgd}
and at 250\,GeV and 500\,GeV in full ILD simulation~\cite{Wang:2020lkq}, and for CLIC at 380\,GeV and 1.5\,TeV~\cite{Mekala:2020zys}. The code of the ILD analysis is available at \url{https://github.com/ILDAnaSoft/ILDbench_extraH}.
New scalar production in the exotic decays of the 125\,GeV Higgs boson was also considered for different scalar decay channels \cite{Liu:2016zki}, however, at generator level only.

\subsection*{Contact \& Further Information}
\begin{itemize}
\item Gitlab wiki: \url{https://gitlab.in2p3.fr/ecfa-study/ECFA-HiggsTopEW-Factories/-/wikis/FocusTopics/EXscalar}
\item Sign up for egroup: ECFA-WHF-FT-EXscalar@cern.ch via \url{http://simba3.web.cern.ch/simba3/SelfSubscription.aspx?groupName=ecfa-whf-ft-exscalar}
\item and/or email the conveners of ECFA WG1 SeaRCHes group:\\ \url{mailto:ecfa-whf-wg1-srch-conveners@cern.ch}
\end{itemize}

\clearpage
\section{{\bfseries LLPs} --- Long-lived particles}
\label{sec:LPPs}
%\section{{\bfseries LLPs} -- Long-lived particles (JL)}
\begin{center}\itshape
%Expert Team: Rebeca Gonzalez Suarez; Juliette Alimena, Jan Hajer, Marcin Kucharczyk, Emma Torro Pastor, Sarah Louise Williams, Aleksander Filip Zarnecki
Expert Team: Juliette Alimena, Rebeca Gonzalez Suarez, Jan Hajer, Marcin Kucharczyk, Emma Torro Pastor, Sarah Louise Williams, Aleksander Filip Zarnecki
\end{center}

Much like the known SM particles, new, BSM particles can also have varying lifetimes, determined by different parameters, such as their mass and couplings. As colliders explore higher energies, heavier and shorter-lived particles become accessible and consequently, collider searches for new particles commonly focus on prompt decays. New long-lived particles (LLPs) however, could provide answers to many open questions in particle physics, such as the nature of dark matter, the origin of neutrino masses, or the baryon asymmetry of the universe.

LLPs have unique experimental signatures, notably displaced vertices, tracks, and jets, but also \emph{broken} tracks, uncommonly high energy loss, or delayed or stopped objects, making them a telltale sign of new physics.
Despite their low background, standard collider techniques often struggle to identify and properly reconstruct LLP signatures.
While the trigger, one of the main issues of LLP searches at high energy colliders, will not necessarily be a problem in a future $e^+e^-$ Higgs/top/EW factory, many other challenges will still be present and will be shared across collider geometry and centre-of-mass energy.

The search for LLPs is taking a central stage in current particle colliders since it offers an exciting experimental alternative and complement to conventional searches for new particles~\cite{Alimena:2019zri,Lee:2018pag}. By planning ahead for this kind of signature, it is possible to optimise the design and performance of future collider experiments to not miss up on the many possibilities LLPs offer~\cite{Bose:2022obr}.

LLP searches form a signature-driven field that can be connected to most BSM models that provide a mechanism to guarantee the longevity of at least one of the new particles.
In the context of $e^+e^-$ colliders~\cite{Blondel:2022qqo, Chrzaszcz:2021nuk}, the possible targets and related topics of interest include but are not limited to: Heavy Neutral Leptons (HNLs), Axion-Like Particles (ALPs), and exotic decays of the Higgs boson.
These topics could attract people already working in this kind of search in current experiments. People working or interested in hardware and software can contribute substantially providing solutions for some of the challenges of this topic. Theorists are needed to clearly defined a set of solid benchmark models. This is in general a good entry point for people interested in new physics searches and detector design. 

\subsection*{Heavy Neutral Leptons}
Sterile neutrinos carry no charge under the SM gauge group allowing in addition to the Dirac also a Majorana mass term.
The heavy mass eigenstate that emerges from the very small mixing with the active SM neutrinos are commonly called HNLs~\cite{Abdullahi:2022jlv}.
Besides the origin of neutrino masses, such models could provide answers to other questions that remain unanswered by the SM, such as the baryon asymmetry in the Universe (BAU), and the particle nature of dark matter.
In order for the SM neutrinos to remain light collider detectable HNLs must appear as pseudo-Dirac pairs of two almost degenerate Majorana fermions.
In such models the mass splitting of the pseudo-Dirac pair governs the light neutrino mass scale and the amount of lepton number violation (LNV) observable in processes involving HNLs.
In addition, leptogenesis is an attractive solution to the question of the origin of matter as it connects the observed BAU with the origin of the light neutrino masses.
In this mechanism, the same HNLs that are responsible for the origin of the light neutrino masses can produce the matter-antimatter asymmetry via their CP-violating decays in the early Universe.
Finally, HNLs can also have connections with dark sectors, such as dark photons and dark scalars, potentially accounting for the observed dark matter.
HNLs have been considered in the context of dark sector models, specifically in fermionic extensions, and can connect to the SM via mixing between sterile neutrinos, dark fermions charged under the new interactions, and the standard neutrinos.
In minimal models, the HNL production and decay are controlled by SM interactions and the mixing between HNLs and the active neutrino and typically result in relatively long lifetimes if the masses are in the MeV--GeV range.

\subsection*{Axions}
Axions were introduced in the 1980s in theories beyond the standard model to address the strong CP problem.
More generally, ALPs~\cite{Bauer:2018uxu} appear in any theory with a spontaneously broken global symmetry and possible ALP masses and couplings to SM particles range over many orders of magnitude.
For large symmetry-breaking scales, the ALP can be a harbinger of a new physics sector at a scale $\Lambda$ that would otherwise be experimentally inaccessible.
Since the leading ALP couplings to SM particles scale as $\Lambda^{-1}$, ALPs become weakly coupled for large new-physics scales.
Accessing the smallest possible couplings, and therefore long lifetimes, is thus crucial to reveal non-trivial information about a whole new physics sector.
In addition, ALPs naturally implement spontaneous electroweak baryogenesis through a cosmic evolution that provides CPT violation~\cite{Im:2021xoy}.
In this scenario, the ALP feebly couples to the Higgs field and gives a small contribution to the Higgs boson mass.

\subsection*{LLPs in Higgs Decays}
A third type of scenario in which LLPs arise is exotic decays of the Higgs boson~\cite{Alipour-Fard:2018lsf}.
The Higgs sector could be extended such that it decays into dark-sector particles, which could be long-lived.
Such signals were first considered in the context of Hidden Valley models, and subsequently found to arise in a variety of well-motivated scenarios including ones involving the electroweak hierarchy problem, models of baryogenesis, and models of neutral naturalness.

\subsection*{Phenomenological and model-building targets}
Due to the plethora of models which can produce LLP signatures, the experimental sensitivities will be derived based on signatures, as sketched below. However these signature-based sensitivities then need to be interpreted and combined in actual BSM models, which might contain LLPs as well as other, promptly decaying new particles -- or new particles which become long-lived only in certain parameter ranges of the model. In particular for the case of a discovery, an important question is what would be needed in order to determine the correct underlying model and its parameters.

In addition, we see interesting opportunities for including projected LLP sensitivities of future colliders in general model-checking tools, for instance the LLP-version of {\sc CheckMATE}~\cite{Desai:2021jsa,Dercks:2016npn}. 

\subsection*{Target physics observables and signatures}
In all cases, the target physics observables comprise the discovery (and exclusion) reach in terms of cross-sections, masses and lifetimes of the LLPs, as well as the precision to which these and other properties of the LLPs can be determined in case of a discovery.  

A large variety of non-mainstream signatures can be considered: 
In the tracking system, Uncommon energy loss patterns in $dE/dx$ as well as displaced tracks and vertices can be pursued, or even ``disappearing tracks'' (in silicon trackers, in continuous tracking even the mini-curler of very soft decay products could be seen).
Non-pointing and/or delayed photons will be of interest in the calorimeters, and in these cases, calorimeter timing would be important. Likewise, non-standard jets, such as emerging jets, trackless jets, or jets with unconventional energy distributions in the calorimeters can be explored. Furthermore, jets with out-of-time decays, such as in later or empty bunch crossings, can be a signature of slowly-moving or stopped particles with a long lifetime.
In the muon system, displaced particles forming a vertex is a signature worth pursuing.
In addition, boosted neutral LLPs could give rise to pairs of collimated muons with no tracks in the inner detector.
Finally, unusual time-of-flight measurements in the muon system and/or the calorimeters could be targeted as well.

\subsection*{Target methods to be developed}

Following on the previous list of signatures, there are a few core methods to be developed for this focus topic. First, the reconstruction of displaced tracks and vertices, both in the inner detector and the muon spectrometers, must be well established. Then, tracking algorithms should be able to reconstruct anomalous $dE/dx$ patterns. Timing capabilities in the calorimeters and for tracks should be developed. Jets must be well-reconstructed and jet taggers explored. For all of these, basic algorithms exist in {\sc Key4hep}, in particular via the {\sc MarlinWrapper} functionality. However they all need significant development to fully explore the physics potential in the area of LLPs. Conceptual, {\sc Delphes}-based studies of displaced vertices and tracks and, to a lesser extent displaced photons are described in Ref.~\cite{Blondel:2022qqo}.

Finally, the estimation of unusual backgrounds must be developed. These backgrounds include instrumental backgrounds such as beam-induced background, pileup, and cavern noise, as well as backgrounds from cosmic-ray muons.

%From this list, work has started and/or is ongoing in the reconstruction and identification of displaced vertices and tracks~\cite{Blondel:2022qqo} and, to a lesser extent in the reconstruction of displaced photons for ALP searches~\cite{Blondel:2022qqo}.

\subsection*{Target detector performance aspects}

This focus topic set requirements on all subsystems of collider detectors, with an emphasis on tracking, timing, and calorimetry. In particular, gaseous main trackers  have advantages in specific energy loss measurements and  pattern recognition which might be difficult to compensate with all-silicon trackers. Analogously, the timing capabilities and the granularity of the chosen ECAL technology might have significant impact on the ability to reconstruct displaced photons and out-of-time signatures. 

Additional experiments following those proposed or running at the LHC and HL-LHC should also be considered, from additional detectors on and off-axis to beam dump experiments, depending on the considered facility.

\subsection*{MC samples needed}
This focus topic will require a number of dedicated signal samples at all centre-of-mass energies. The correct interfacing of late decays from the MC generators to the detector simulation in {\sc Geant4} is non-trivial, but working examples exist~\cite{Schnoor:llpgen}. Specifically for this focus topic, additional  simulations of accelerator-related backgrounds (beamstrahlung, halo muons, beam gas interactions) might be needed.

Specifically at the $Z$ pole, high statistics of \eg $Z\rightarrow b\bar{b} \tau\tau$ will be needed, as well as dedicated filtering strategies to enhance tails of distributions where displacement can be found.

\subsection*{Existing tools / examples}

Work on LLPs is ongoing based on fast simulation of IDEA~\cite{Blondel:2022qqo} and full simulation of ILD~\cite{Klamka:2023kmi}.
An FCC LLP code tutorial already exists~\cite{LLPtutorial}. The reconstruction of displaced photons from ``invisible $\to$ invisible + $\gamma$'' as well as the prospects for measuring key properties of the two invisible particles have been studied in full simulation of ILD~\cite{Wattimena:2010zz}.

\subsection*{Contact \& Further Information}
\begin{itemize}
\item Gitlab wiki: \url{https://gitlab.in2p3.fr/ecfa-study/ECFA-HiggsTopEW-Factories/-/wikis/FocusTopics/LLPs}
\item Sign up for egroup: ECFA-WHF-FT-LLP@cern.ch via \url{http://simba3.web.cern.ch/simba3/SelfSubscription.aspx?groupName=ecfa-whf-ft-llp}
\item and/or email the conveners of ECFA WG1 SeaRCHes group:\\ \url{mailto:ecfa-whf-wg1-srch-conveners@cern.ch}
\end{itemize}

\clearpage
\section{{\bfseries EXtt} --- Exotic top decays}
\label{sec:EXtt}
% mainfile: ../benchmarks

\begin{center}
{\itshape
Expert Team:
Nuno Castro,
Marina Cobal,
Gauthier Durieux,
Roberto Franceschini,
Mar\'ia Teresa N\'u\~nez Pardo de Vera,
Kirill Skovpen,
Marcel Vos
}
\end{center}

So far the study of top quark physics at the $e^{+}e^{-}$ factory has concentrated mostly on an effective field theory (or anomalous couplings) encoding the microscopic details of new physics characterised by masses larger than the energy scales experimentally accessible.
These include modified electroweak interactions and flavour-conserving contact interactions as well as flavour-changing neutral currents (FCNCs) which are negligibly small in the standard model.

FCNC interactions arising at dimension six in the effective field theory of the standard model (SMEFT) include: $tqg$, $tq\gamma$, $tqZ$, $tqH$, $tq\ell\ell^{(\prime)}$.
These interactions can be probed both in top-quark decay and (associated) production, at hadron and lepton colliders.
Top-quark FCNC decays are not foreseen in the SM at any observable level, e.g.\ $t\to c h$ is predicted in the SM to be at $\text{BR}=3\cdot 10^{-15}$. The FCNC BR can reach observable level in BSM. However, the potential of new lepton colliders on these FCNC interactions may be limited in common models, 
because the particles (e.g.\ squarks of SUSY) which could produce observable BRs also have plenty of other observable consequences, which are already subject to significant constraints.
%\cite{extt:ongoingDtudy}

A new direction of research for BSM related to the top quark is  the study of decays into new-physics particles. These decays are foreseen in well motivated models of new physics, experience less tight experimental constraints from other searches than decays in SM states, and have received little attention so far at both $e^{+}e^{-}$ and hadron machines. 

Following the proposal of Refs.~\cite{Banerjee:2018fsx,Bahl:2023aa} an interesting example for
  BSM decays of the top quark into new scalars and light quark
flavours is 
\begin{equation}
t \to \phi q,\quad \phi \to b\bar{b}\,. \label{eq:top-decay-phi}
\end{equation}
This production in top decays is particularly interesting for the case that the $\phi$ has so strongly reduced couplings to the $Z$ boson that it would escape detection in searches for associated $Z\phi$ production (``scalar strahlung'').

Current results from LHC show a potential to probe BR in these exotic decays modes of order $10^{-4}$. For example a recent search for the case $\phi\to b\bar{b}$ and $q=u,c$ using 139~$\text{fb}^{-1}$ found an upper limit on the BR around
the few $10^{-4}$ ballpark for $20\text{ GeV}<m_{\phi}<160\text{ GeV}$~\cite{ATLAS:2023mcc}.
This leaves plenty of room for improvement for $e^{+}e^{-}$ machines, that could reach up to $O(10^{-6})$ for particularly clean signals --- or for precise property determinations in case HL-LHC should discover such a decay in the future. In addition, the LHC has difficulties to search for too light new physics, whereas trigger-less $e^{+}e^{-}$ machines could much more easily attack such signals from light new physics appearing in top quark decays.

The topics proposed in the following are particularly well suited as ``low barrier'' projects that could be undertaken e.g.\ by colleagues working on top quark physics at the LHC.

\subsection*{Theory and phenomenological targets \label{sec:extt-pheno-targets}}

Most $t \to \phi q$ searches so far have focused on $\phi=h$, for instance at CLIC~\cite{Zarnecki:2018wde}. However it is highly motivated to consider other scalars than the SM Higgs boson $h$. In particular, the possibility to consider scalars other than the Higgs boson frees up from constraints from other observables related to the properties of the Higgs boson and other tests of the SM. In any case a ``generic scalar'' search will encompass the Higgs boson as a sub-case, thus allows to obtain results for the SM Higgs at little or no cost at all.

From this starting point the {\bf main phenomenological target} is to quantify the discovery reach of the HET factory below the HL-LHC limits at BR$(t\to\phi q)\lesssim10^{-4}$ in the channel $\phi\to bb$  in the mass range $m_{\phi}\in [10,172]$~GeV, and to study the obtainable precisions on mass and other properties of the $\phi$.

In addition, the search in other decay modes of $\phi$, e.g.\  $\phi\to\gamma\gamma, c\bar{c}, s\bar{s}, gg,\,\text{and invisible}\,,$ especially for $m_{\phi}<10\text{ GeV}$, which are also very challenging for LHC. Each of this can provide a  new target for detector performances and reconstruction methods.

\subsection*{Detector Performances and Methods} 
The search for the exotic top decay eq.~(\ref{eq:top-decay-phi}) hinges around reconstruction performances for multi-jet events. A key development concerns methods for $b$ tagging for a kinematics different from the typical $t \to b W$ and $Z\to b\bar{b}$, in particular for ``soft'' and/or ``small-opening-angle'' $b$-jets from a light $\phi$.   
These methods can be explored for $\phi$ search in top decay, but also for a number of other scenarios where light sources of heavy flavours emerge, e.g. $h\to \phi_{1} \phi_{2}$ of interest in the {\bf EXscalar} Focus Topic (Sec.~\ref{sec:EXscalar}).

Considering the fact that each event will contain two top quarks, there will be up to six jets in the event, which can lead to significant jet clustering errors. Thus the development of jet algorithms and kinematic fitting will be essential for a realistic performance estimate.

Extending the scope from $\phi \to b\bar{b}$ to $\phi \to c\bar{c}, s\bar{s}, \gamma\gamma, \text{invisible}$ will add the need for corresponding tagging algorithms, especially for the lower-mass cases.

\subsection*{MC sample needs}
For the generation of signal event, UFO models with all the possible decay modes for $\phi$ are available:
    \begin{itemize}
        \item for FCNC at NLO in QCD, where changing the Higgs mass would make it look like a $\phi$ at \url{http://feynrules.irmp.ucl.ac.be/wiki/TopFCNC}.
        \item similarly, LO SMEFT models with all the relevant top-quark interactions can be found at \url{https://feynrules.irmp.ucl.ac.be/wiki/dim6top} and \url{https://smeftsim.github.io}.
    \end{itemize}
    
In principle more refined studies may include inclusive $t\bar{t}$,$t\bar{t}V$, multi-boson
background generation.

\subsection*{Contact \& Further Information}
\begin{itemize}
\item Gitlab wiki: \url{https://gitlab.in2p3.fr/ecfa-study/ECFA-HiggsTopEW-Factories/-/wikis/FocusTopics/EXtt}
\item Sign up for egroup: ECFA-WHF-FT-EXTT@cern.ch via \url{http://simba3.web.cern.ch/simba3/SelfSubscription.aspx?groupName=ecfa-whf-ft-extt}
\item and/or email the conveners of ECFA WG1 SeaRCHes group:\\ \url{mailto:ecfa-whf-wg1-srch-conveners@cern.ch}
\end{itemize}

\clearpage
\section{{\bfseries CKMWW} --- CKM matrix elements from $W$ decays} 
\label{sec:CKMWW}
%\section{{\bfseries CKMWW} -- CKM matrix elements with on-shell and boosted $W$ decays at $\sqrt{s} \ge m_W$ (FM)} 

\begin{center}
{\itshape
Expert Team:
Marzia Bordone,
Ulrich Einhaus,
Pablo Goldenzweig,
Zoltan Ligeti,
David Marzocca,
St\'ephane Monteil,
Michele Selvaggi
}
\end{center}

%\centerline{\bf\color{red}{There is a separate temporary overleaf document, so don't edit this section for now.}}

\subsection*{Introduction}

On-shell hadronic $W$ decays offer the possibility to directly measure CKM matrix elements in a way completely independent from low-energy measurements, provided a sufficiently good flavour-tagging efficiency is achievable.
An $e^+e^-$ Higgs/EW/top factory, with the possibility  of recording several $10^{8}$ $W$ events in a clean environment, is the ideal machine for such measurements.

In principle, the magnitude of all CKM matrix elements $|V_{ij}|$, except for those involving the top quark, can potentially be measured in this way from the decay $W^- \to \overline{u}_i d_j$ (where $u_i = (u, \, c)$ and $d_j = (d, \, s, \, b)$). The main goal of this Focus Topic is to study the prospects for $|V_{cb}|$ and $|V_{cs}|$, where unprecedented precision could be reachable. However, we will also consider the other matrix elements, comparing the reach with the state-of-the-art precision from low energy probes.

\subsection*{Motivation and phenomenological targets}

The knowledge of the magnitudes $|V_{cs}|$ and $|V_{cb}|$ from $W$ decays provides a consistency check of the unitarity of the CKM matrix. The comparison of the $|V_{cs}|$ magnitude to its indirect determinations inferred from leptonic charmed meson decays is a direct test of the SM decay constant parameters.
The measurement of $|V_{cb}|$ with on-shell and boosted $W$ decays may shed light on the longstanding discrepancy observed in exclusive and inclusive determinations of $|V_{cb}|$ obtained from semileptonic decays of $b$-flavoured particles~\cite{HFLAV21}. Furthermore, the  $|V_{cb}|$ magnitude controls the normalisation of the unitarity triangle.
A quasi-model-independent global analysis of neutral $B$-meson observables will benefit from an improved precision at the horizon of 2040 \cite{Charles:2020dfl}.          

While the uncertainty in $|V_{cb}|$ is already systematically dominated and can only be improved via a better understanding of detector performance, the expected precision in $|V_{ub}|$ from Belle II is approximately 1.3\% for 50~ab$^{-1}$ of integrated luminosity, assuming a factor of 5 improvement in lattice QCD uncertainties \cite{Belle-II:2018jsg}.
From this perspective, measuring CKM matrix elements directly from $W$ decays, which do not require lattice inputs, could also provide a way to benchmark different lattice QCD computations.

\subsection*{Target physics observables}

A starting point for computing the achievable precision on each CKM matrix element is the total number of $W$ boson produced, which can then decay to each specific hadronic channel.
The total hadronic branching ratio of a $W$ boson is \cite{Workman:2022ynf}
\begin{equation}
    {\rm BR}(W^- \to \text{hadrons}) = (67.41 \pm 0.27)\% ~.
\end{equation}
The branching ratio for each quark channel can be approximated by (we neglect quark mass effects)  ${\rm BR}(W^- \to \bar{u}_i d_j) \approx \frac{1}{2} |V_{ij}|^2 {\rm BR}(W^- \to \text{hadrons})$, with the result shown in Table~\ref{table:WquarkBrs}.
In the bottom row we show the number of events in each channel assuming a total number $N_W = 10^8$ of $W^\pm$ pairs produced and the lower limit for the statistical uncertainty in $|V_{ij}|$ assuming an unrealistic reconstruction efficiency of 100\%, computed as $\delta^{\rm th}_{V_{ij}} = \frac{1}{2} N_{\text{ev}}^{-1/2}$.

\begin{table}[htb]
\centering
\caption{Standard Model branching ratios of the $W^-$ boson into quarks and the corresponding number of events, in each channel, assuming a number $N_W = 10^8$ of $W^\pm$ of boson pairs produced (we include the charge conjugate channel in the counting). The last row shows the theoretical lower limit on the statistical uncertainty $\delta_{V_{ij}}$ in each CKM matrix element assuming 100\% reconstruction efficiency and $10^8$ $W$ pairs produced.}
\begin{tabular}{ | c | c | c | c | c | c | c | }
\hline
$W^- \to $ & $\bar{u} d$ & $\bar{u} s$ & $\bar{u} b$ & $\bar{c} d$ & $\bar{c} s$ & $\bar{c} b$ \\ \hline
${\rm BR}$ & 31.8\% & 1.7\% & $4.5 \times 10^{-6}$ & 1.7\% & 31.7\% & $5.9 \times 10^{-4}$ \\ \hline
$N_{\rm ev}$ & $64 \times 10^6$ & $3.4 \times 10^6$ & $900$ & $3.4 \times 10^6$ & $63 \times 10^6$ & $118 \times 10^3$ \\ \hline
$\delta^{\rm th}_{V_{ij}}$ & $0.0063$ \% & 0.027 \% & 1.7 \% & 0.027 \% & 0.0063 \% & 0.15 \% \\ \hline
\end{tabular}
\label{table:WquarkBrs}
\end{table}

A study of the particle reconstruction capabilities of each detector design will allow to derive reconstruction efficiencies for each decay channel, enabling an evaluation of the possible reach for the CKM matrix elements.
A first estimate for the sensitivity reach in $V_{cb}$ can be found in Ref.~\cite{MHS}, where a precision $\delta_{V_{cb}} \approx 0.4\%$ is projected for $10^8$ $W$ pairs, a substantial improvement on the expected precision from $B$ meson decays.

\subsection*{Target methods to be developed}

Particle-identification performance of the detector is the crucial characteristic to be studied and potentially improved to estimate the quark-jet flavour tagging efficiency. Tagging efficiency for $s$, $c$, and $b$-jets are relevant for $|V_{cb}|$ and $|V_{cs}|$, while also light quark tagging and gluon-jet rejection become relevant for CKM elements of light quarks.
Time-of-flight PID seems to be feasible for any future collider detector, but it provides PID only at low momenta ($\pi/K$ up to about 5 GeV, $K/p$ up to about 10 GeV with currently achievable several 10 ps timing resolution). While this is crucial for flavour physics at the $Z$ pole, its impact to tag leading particles in jets at the $W$ threshold or the 240/250 GeV Higgs is limited; above that energy it becomes very small.
PID with gaseous trackers (time projection chamber or drift chamber with d$E$/d$x$ or d$N$/d$x$) can provide $\pi/K$ separation up to 20 to 40 GeV, depending on the system (effective depth, granularity, available counting method).
The PID performance of a dedicated Cherenkov detector promises the highest momentum reach of 50 GeV or above and shall be studied as well.

Over the years, {\sc LCFIPlus}~\cite{LCFIPlus} has become a benchmark and default flavour tagger for future $e^+e^-$ collider studies. It uses a boosted decision tree to tag $b$ ,$c$ and ``other'' jets.
Currently, a number of novel approaches to flavour tagging are being worked on, using neural networks and allowing for a more comprehensive tagging of all quark flavours. For an overview see Ref.~\cite{ECFARecoOverview2023}.
An important next step is the implementation of these taggers in {\sc Key4hep} in order to make them available to the full community.

A central question of this study is to determine the dependence of the CKM matrix element measurement precision on the flavour tagging performance, as well as the dependence of the flavour tagging on the PID performance. This will allow to optimise the flavour tagger and potentially also the overall detector design with respect to PID-sensitive subdetectors.
Furthermore, the precise needs for hadronic particle identification ($p / K / \pi$) shall be assessed.

\subsection*{MC samples needed}

The ILD concept group has produced large full-simulation MC samples to study physics benchmarks, including $WW$ production.
At 500 GeV, about 500 fb$^{-1}$ are available \cite{ILD_MCProd_2018}. The more recent 250 GeV MC production contains 5 ab$^{-1}$ for either opposite-sign 100\%-polarisation combination $(+-, -+)$ and 1 ab$^{-1}$ for either like-sign 100\%-polarisation combination $(--, ++)$ \cite{ILD_MCProd_2020}, exceeding the integrated luminosity of the proposed data taking by more than a factor of 3. Flavour tagging was not part of these productions, but {\sc LCFIPlus} is typically added as default tagger on the `analysis-level.'

\noindent On the FCC-ee side, full simulation for CLD and IDEA detectors with several $10^{^8}$ $W$ decays are needed.
% From Uli: Do we have numbers for FCC-ee detectors?

\subsection*{Contact \& Further Information}
\begin{itemize}
\item Gitlab wiki: \url{https://gitlab.in2p3.fr/ecfa-study/ECFA-HiggsTopEW-Factories/-/wikis/FocusTopics/CKMWW}
\item Sign up for egroup: ECFA-WHF-FT-CKMWW@cern.ch via \url{http://simba3.web.cern.ch/simba3/SelfSubscription.aspx?groupName=ecfa-whf-ft-CKMWW}
\item and/or email the conveners of ECFA WG1 FLAVour group:\\ \url{mailto:ecfa-whf-wg1-flav-conveners@cern.ch}
\end{itemize}

\clearpage
\section {{\bfseries BKtautau} --- $B^0 \to K^{0*} \tau^+ \tau^-$}
\label{sec:BKtautau}
%\section {{\bfseries BKtautau} -- $B^0 \to K^{0*} \tau^+ \tau^-$ (FM)}

%-------------------------------------------------
\begin{center}
{\itshape
Expert Team:
Paula Collins,
Pablo Goldenzweig,
Jernej Kamenik,
Matt Kenzie,
Elisa Manoni,
David Marzocca,
Tristan Miralles,
St\'ephane Monteil,
Fabrizio Palla,
Luiz Vale Silva,
Aidan Wiederhold
}
\end{center}

\subsection*{Theoretical and phenomenological targets}

This process tests the partonic FCNC transition $b \to s \tau^+ \tau^-$. Present limits on this are very weak: $\text{BR}(B^+ \to K^+ \tau^+ \tau^-) < 2.25 \times 10^{-3}$ at 90\% CL \cite{BaBar:2016wgb}; $\text{BR}(B^0 \to K^{*0} \tau^+ \tau^-) < 3.1 \times 10^{-3}$ \cite{Belle:2021ecr} at 90\% CL; and $\text{BR}(B_s^0 \to \tau^+ \tau^-) < 6.8 \times 10^{-3}$ at 95\% CL \cite{LHCb:2017myy}. Belle II and LHCb are expected to push these limits to the $10^{-4}$ ($10^{-5}$) level for the leptonic (semi-leptonic) mode. These should be compared with the SM prediction, which is at the $\sim 10^{-7}$ level for $\text{BR}(B \to K^{(*)} \tau^+ \tau^-)$. No limits are currently available for the $B_s^0 \to \phi \, \tau^+ \tau^-$ mode \cite{Kamenik:2017ghi}.

The large number of boosted $B$ mesons produced from $Z$ decays at a Tera-$Z$ factory may  substantially improve on the expected Belle II limits and possibly reach the sensitivity necessary to test the SM prediction.

The theoretical interest in this decay mode is large, since it would allow to test a FCNC process involving third generation leptons, complementing the present precise measurements of $b \to s \ell^+ \ell^-$ with light leptons.
As a specific example, large effects due to New Physics can be expected in connection with the $R(D^{(*)})$ anomaly. These involve the $b \to c \tau^- \bar{\nu}_\tau$ transition, which in turn can be related to $b \to s \tau^+ \tau^-$ via $SU(2)_L$ invariance and flavour symmetries \cite{Greljo:2015mma,Capdevila:2017iqn}. The same symmetries relate to $b \to s \nu_\tau \bar{\nu}_\tau$ transitions, which can be tested, for instance, in $B \to K^{(*)} \nu \bar{\nu}$ decays.

 This motivates the inclusion of the so-called golden channels $B \to K^{(*)} \nu \bar{\nu}$ in this Focus Topic. These are not affected by long-distance charm loop effects, allowing for lower theoretical uncertainties.
Belle II recently performed a search for $B^+ \to K^+ \nu \bar{\nu}$ decays in a $362~{\rm fb}^{-1}$ data set and determined the branching fraction to be $(2.3 \pm 0.5 ^{+0.5}_{-0.4}) \times 10^{-5}$ \cite{Belle-II:2023esi}.
This measurement has a significance of 3.5$\sigma$ and is the first evidence for the decay. A combination with previous measurements gives an average of $(1.3 \pm 0.4) \times 10^{-5}$ resulting in a 2.1$\sigma$ tension with the SM prediction of $(4.44 \pm 0.30) \times 10^{-6}$ \cite{Becirevic:2023aov}.
A recent study has been performed to give a baseline estimate for the sensitivity achievable at FCC-ee~\cite{Amhis:2023mpj}, including studies on the particle identification and vertex resolution requirements.
The FCC-ee will be able to substantially improve upon Belle II measurements of $b \to s \nu \bar{\nu}$ decays by obtaining branching fraction measurements with $\mathcal{O}(1\%)$ sensitivity. This will likely allow for differential branching fraction measurements in these modes and will enable measurements of $B_s^0$ and $\Lambda_b^0$ decays which are impossible at Belle II.

%-------------------------------------------------
\subsection*{Target physics observables}

The $B^0 \to K^{*0} \tau^+ \tau^-$ decay was considered in Ref.~\cite{Kamenik:2017ghi} as a mode to test the $b \to s \tau^+ \tau^-$ transition.
While the two neutrinos emitted from $\tau$ decays complicate the experimental search, an excellent reconstruction of all the vertices,  considering hadronic $\tau$ decays, can help to close the kinematics of the process.
Ref.~\cite{Kamenik:2017ghi} also considers $\tau$ polarisation as a further observable to disentangle the Standard Model from possible New Physics contribution.

To study the $b \to s \nu \bar{\nu}$ transition, branching fraction measurements of $B^0 \to K_S^0 \nu \bar{\nu}$, $B_s^0 \to \phi \nu \bar{\nu}$, $B^0 \to K^{*0} \nu \bar{\nu}$ and $\Lambda_b^0 \to \Lambda \nu \bar{\nu}$ decays have been considered.
Given the interest in the Belle II measurement, the $B^+ \to K^+ \nu \bar{\nu}$ decays should also be considered, although this is experimentally more challenging due to the lack of a secondary decay vertex.

%-------------------------------------------------
\subsection*{Target methods to be developed}

\begin{itemize}
 %   \item At hand but to be refined: topological reconstruction.  
 %   \item Secondary vertex fitting algorithm.  
 %\item topological reconstruction, dependency on $\tau$ and $K^*$ reconstruction, vertexing

    \item Topological reconstruction of the decay $B^0 \to K^{*0} \tau^+ \tau^-$:  the proof-of-principle of the kinematic fits that determine the missing neutrino momentum for three-prongs $\tau$ decays in $b \to s \tau^+\tau^-$ has been established in Refs.~\cite{Kamenik:2017ghi,Li:2020bvr}.  The reconstruction technique relies on the experimental knowledge of the production and decay vertices of both the $b$ hadron and the $\tau$ leptons. The vector between the production and decay of each particle provides its momentum direction and hence fixes two degrees of freedom of the problem. The knowledge of the momentum directions of the $b$ hadron and the two $\tau$ leptons, complemented with the known $\tau$ mass, provide the six constraints necessary to determine the two missing neutrino momentum vectors. Further developments of this method are desirable and can be obtained, for instance, by introducing kinematic fits of the decay chain accounting for the actual vertex resolutions such as the {\sc DecayTreeFitter} algorithm~\cite{Hulsbergen:2005pu} developed for the BaBar experiment and presently used in the LHCb and Belle II experiments.  

    \item Three-prong $\tau$ decays account for approximately $10\%$ of the decay width. It is therefore relevant and desirable to explore novel methods to experimentally reconstruct  the leptonic $\tau$ decays, for at least one of the two $\tau$ decays. In contrast to three-prong decays, the leptonic decay kinematics cannot be fully closed with the experimental vertex measurements. The global topological information on the decay could, however, help to select further events to add up to the global statistical significance.  

    \item The decay vertex fitting and the selection of the final states of interest: the secondary vertex fitting is a common tool for all final states of interest in this Section. The transitions $b \to s \tau^+\tau^-$ and  $b \to s \nu\bar{\nu}$ would benefit from both an excellent secondary vertex resolution and reconstruction efficiency, as a starting point to their selection. The Tera-$Z$ factories machines feature a very well defined luminous region, at least in the transverse plane ($\{ \sigma_x, \sigma_y \} =  \{ 3 , 0.02 \} \; \mu{\rm m}$), that can be used to optimise the set of tracks which are inconsistent with the primary vertex. This would be particularly instrumental for high track-multiplicity final states.   

\end{itemize}

%-------------------------------------------------
\subsection*{Target detector performance aspects}

Precise vertexing is essential to most of the heavy-flavour measurements: time-dependent $B^0_s$ CP asymmetries studies, for instance, require to resolve the $B^0_s$ proper-time significantly better than the $\sim$350\,fs oscillation period of the $B^0_s$ meson. State-of-the-art vertex detectors envisaged for a Tera-$Z$ factory are likely to fulfil this requirement, with resolutions of a few microns for multi-track vertices~\cite{FCC:2018evy}. The reconstruction of the decay modes of interest in this Section impose more challenging detector requirements and would benefit from even better performance. Various improvements can be envisaged and tested with these modes as a case in point. These include, but are not limited to:
%Let's single out few of them to illustrate the targets of this study:

\begin{itemize}  
    
    \item shortening the distance of the first tracking layer to the interaction point (beam-pipe radius); 
    \item reducing the overall material in the detector, as well as assessing the cooling system, and the use of bent sensors;  
    \item understanding the impact of improving hit resolution. 

\end{itemize} 

%-------------------------------------------------
\subsection*{MC samples needed}
\begin{itemize}
    \item At hand: fast simulation with a state-of-the-art tracker. Emulation of an arbitrarily good vertex detector. 
    \item Fast simulations featuring actual vertex detector(s) implementations. 
    \item Full simulation studies.   
\end{itemize}

%-------------------------------------------------
\subsection*{Existing tools / examples}

Work on the physics reach of the modes $B^0 \to K^{*0} \tau^+ \tau^-$ and $B^+ \to K^{+} \nu\bar{\nu} $ is ongoing based on fast simulation of the IDEA concept detector. The detailed simulation of the signals and the backgrounds relies on the {\tt EvtGen} generator, interfaced with {\tt Pythia}. A centralised simulation is available.  Selections have been designed in both cases and can serve as starting points to study these decay modes or add up further relevant companion modes.  

%-------------------------------------------------
\subsection*{Contact \& Further Information}
\begin{itemize}
\item Gitlab wiki: \url{https://gitlab.in2p3.fr/ecfa-study/ECFA-HiggsTopEW-Factories/-/wikis/FocusTopics/BKtautau}
\item Sign up for egroup: ECFA-WHF-FT-BKTAUTAU@cern.ch via \url{http://simba3.web.cern.ch/simba3/SelfSubscription.aspx?groupName=ecfa-whf-ft-BKtautau}
\item and/or email the conveners of ECFA WG1 FLAVour group:\\ \url{mailto:ecfa-whf-wg1-flav-conveners@cern.ch}
\end{itemize}

\clearpage
\section{{\bfseries TwoF} --- EW precision: 2-fermion final states ($\sqrt{s}=M_Z$ and beyond)}
\label{sec:TwoF}
%\section{{\bfseries TwoF} -- EW precision: 2-fermion final states ($\sqrt{s}=M_Z$ and beyond) (FM)}
\begin{center}
{\itshape
Expert Team:
%Emanuele Bagnaschi, Freya Blekman, Adrian Irles, Daniel Jeans, Eram Rizvi, Alessandro Vicini
Emanuele Bagnaschi, Adri\'an Irles, Daniel Jeans, Alessandro Vicini
}
\end{center}
\subsection*{Introduction}

%TODO: add reference to the "estimation" of future measurements for ILC

The precision of the determination of the EW couplings of gauge bosons to fermions is expected to improve by several orders of magnitude at future $e^{+}e^{-}$ colliders \cite{FCC:2018evy} with respect to the legacy measurements from LEP and SLC\cite{ALEPH:2005ab}.
Such precision will be achievable thanks to the higher luminosities, longitudinally polarised beams (in the case of linear colliders), a wider range of collider energies, precise modern detectors with improved reconstruction, and improved theoretical modelling.

The unprecedented statistical power provided by future colliders will require a large effort on the control and understanding of systematic uncertainties from theory and experiment. Indeed, the run at the $Z$ pole foreseen by FCC-ee will offer 500 times smaller statistical uncertainties than those of previous measurements~\cite{FCC:2018evy}. %A similarly high precision will be reached 
A significant improvement in precision could also be reached 
at the ILC~\cite{ILCInternationalDevelopmentTeam:2022izu}. %\JB{I modified this sentence because the previous sentence refers to statistical uncertainties, and here it is $10^{12}$ at FCCee vs $10^9$ at ILC, so it is not correct to say similar precision, if referring to stat uncertainties (though in the end $Z$ pole measurements are sys limited so the statement is not very off)}
This requires remarkably stable operation of the detectors and accelerators. All of these elements are indispensable to achieve the highest level of scientific output from specialised physics runs conducted at the $Z$ pole.

The LEP and SLC colliders probed the gauge structure of the SM at the quantum level, finding an overall good agreement with theory predictions. However, some tensions in the determination of the weak effective mixing angle for different flavours are still unresolved. Future colliders will play an important role in clarifying these issues.
Moreover, the remarkable precision of the experimental measurements, coupled with refined theory predictions, will test the SM up to a new degree.
Any observed deviation will be interpreted as an indirect sign of new physics.
Moreover, for the investigation of the Higgs sector and for the searches for new physics at higher energies, more precise determinations of the EW couplings to fermions are required \cite{Ellis:2020unq}. 

Projections for the determination of the electroweak couplings of the $Z$ boson to fermions from measurements at a future $e^+e^-$ collider running at the $Z$ pole are available in Refs.~\cite{ILCInternationalDevelopmentTeam:2022izu,FCCTeraZPaus}.
These studies cover tests of lepton universality and the couplings to heavy quarks.
More work is required to exploit final states involving light quark families, for instance using strange-tagging techniques. 

Final states with two fermions will also be studied at runs with higher energies.
These data will be more difficult to interpret in terms of couplings
due to the impossibility of isolating the $Z$-exchange contribution. 
However, they give the interesting possibility of studying the energy dependence of the EW couplings.
Moreover, the mechanism of radiative return allows to study the invariant mass distribution of the fermion pair,
including the $Z$ pole region, also at beam energies much larger than half the $Z$ mass. To exploit this opportunity, the development of new experimental and theoretical techniques has been started~\cite{Mizuno:2022xuk}.
% TOOD: add reference

\subsection*{Theoretical and phenomenological aspects}

The cross section for fermion pair production at the $Z$ peak will be measured with a $\mathcal{O}(10^{-4})$ relative precision at future $e^+e^-$ colliders. Following the approach that was chosen at the time of LEP and SLC colliders it is possible to parameterise the $Z$ resonance in terms of pseudo-observables. To match the experimental precision, these have to be computed including three-loop electroweak corrections~\cite{Blondel:2019qlh,Chen:2020xzx}, which are not fully available today. The relative statistical precision around the $Z$ peak and at higher energies will be better than $\mathcal{O}(10^{-3})$. To exploit these data, predictions for cross sections and asymmetries at NNLO electroweak will be required. More work in this direction is needed from the theory side. As an intermediate step, NNLO mixed QCD-EW corrections to $e^+e^- \to q\bar{q}$ could be studied using the results of Refs.~\cite{Bonciani:2021zzf,Armadillo:2022bgm}. In the absence of beam polarisation, the determination of the $A_e$ asymmetry hinges upon the measurement of $\tau$ polarisation, which can provide additional information and redundancy in the presence of beam polarisation. Another approach would be to directly fit the effective leptonic weak mixing angle from the physical observables, cross sections and asymmetries, keeping it among
the input parameters of the models~\cite{Chiesa:2019nqb}. Flavour identification of final states with hadronic jets will benefit from recent work developed for the LHC~\cite{Czakon:2022wam,Gauld:2022lem,Caola:2023wpj}.

The two-fermion final-state inclusive cross section has a well defined energy dependence in the SM. Deviations from this prediction would be a sign of new physics. This can be interpreted in terms of higher-dimensional operators, or in terms of explicit models of light new physics. In the case of an EFT interpretation, the effect of higher dimensional operators needs to be computed in a consistent way together with the SM predictions, as for instance it was done recently for the Drell-Yan process at NLO~\cite{Dawson:2021ofa}.
An interesting target for BSM exploration, given the current observed tension with the SM prediction, would be $A_{\rm FB}^b$. 
The measurement of final states with third generation fermions, are also especially interesting due to the role of the third family in BSM models. Another example in that direction is the study of $\tau$ spin correlations. 

\subsection*{Target physics observables}
\begin{itemize}
    \item Total and differential cross sections as well as asymmetries at different energies for two-fermion inclusive final states; in the case of $\tau$ leptons, including the $\tau$ polarisation.
    \item Combination of $Z$-pole and off-pole measurements to study the energy dependence of EW interactions: separate EW couplings and four-fermion interactions.
    \item Final states with one isolated $\gamma$, interpreted as $Z$ decay to neutrino, or as a new physical signal
    \item $\tau$ spin correlations for CP violation, EDM, $a_\tau$, Bell's inequality, at the $Z$ pole and at higher energies. 
\end{itemize}

\subsection*{Experimental aspects}

\subsubsection*{Full simulation studies}

Several full simulation studies have been conducted or are ongoing on $e^{+}e^{-}$ collisions producing $b\bar{b}$, $c \bar{c}$, $s \bar{s}$, and $\tau^+ \tau^-$ final states. 
Most of these studies are based on high-energy collision (250 GeV, 500 GeV) simulations performed by the ILD concept group using the ILC beam conditions and ILD model.
Ref.~\cite{Jeans:2019brt} reports the work on the $\tau^{+}\tau^{-}$ reconstruction at 500 GeV for different decay modes and the potential of measuring the $\tau$ polarisation.
Several studies have been conducted using final state $q\bar{q}$ production \cite{2fPaestum} and $b\bar{b}$ and $c\bar{c}$, in particular \cite{Irles:2023nee,Irles:2023ojs,Marquez:2023guo,Irles:TBA}.
The work for the $b\bar{b}$ and $c\bar{c}$ includes the study of data-driven techniques (such as double tagging and double charge measurement) for the determination of $R_{q}$ and $A_{\rm FB}^{q}$
reducing the usage of MC simulations for the modelling to the minimum (i.e. for better control of the systematic uncertainties such as fragmentation functions or angular correlations due to QCD corrections).
These techniques could not be maximally exploited in past experiments due to reduced yields and/or reduced vertexing capabilities. In Refs.~\cite{Irles:2023nee,Irles:2023ojs,Marquez:2023guo,Irles:TBA}, 
the potential of using $dE/dx$ or $dN/dx$ for charged hadron identification has also been studied and proven to have a large impact on the final uncertainties, especially for the $c$-quark case, where the kaon identification becomes crucial for the charge reconstruction of the jet. In these studies, the power of the beam polarisation for BSM disentangling and systematic uncertainties control is also studied.
Finally, the power of discrimination of several BSM models in Gauge-Higgs Unification theories was investigated in Ref.~\cite{Irles:TBA}, showing that thanks to the beam polarisation and energy reach of ILC, 
KK resonances of up to $\sim30$ TeV could be probed. This work includes the study of the impact of the assumed level of precision on the $Z$-fermion couplings in the search of BSM physics. 
Dedicated full-simulation studies of experimental prospects for $A_{\rm FB}^{b/c}$ reconstruction at the $Z$ pole have been started \cite{FCCTeraZAFB}. A comprehensive assessment of the systematic uncertainties
that could challenge the superb statistical expectations must be addressed. 
Full-simulation studies for the comparison of the potential of $Z$-pole running and radiative return events at e.g. 250 GeV CM energy are still missing and should be addressed.

In all these works the importance of flavour tagging, particle identification (kaon identification, photon identification) and precise vertexing in the barrel and also the forward region is emphasised.
Some of the detector target performance aspects to be addressed and better understood are listed below.

\subsubsection*{Target detector performance aspects}
\begin{itemize}
    \item Design of inner tracking detector systems and their influence on vertex finding, with a focus on the forward region (the most important region for physics sensitivity).
    \item Charged hadron identification detectors, especially of kaons for $b, c, s$ tagging and charge measurement using $dE/dx$, $dN/dx$, time-of-flight detectors, RICH detectors, etc.
    \item impact of ECAL design (e.g. granularity an energy resolution) on tau decay mode identification.
    \item Impact of high-energy photon identification. This will be needed for measurements in the continuum where the return to the $Z$ pole contributes to the background but not to the signal. These photons tend to be found in the forward regions. 
    \item Hermeticity of detectors: to study $Z$-couplings at 240/250 GeV with radiative return events, we will have to look at events in which the ISR photon has escaped the detector through the beam pipe. Proper estimations of the missing energy (via angular measurements, for example) are required, and these depend on the detector hermeticity.
    \item Typical 2f signatures are back-to-back; detectors often have an even-fold symmetry in $\phi$. Would eliminating back-to-back detector cracks (e.g. by adopting an odd-fold symmetry) help reduce systematic effects?
\end{itemize}

The optimisation of the detector concepts should also address the difference of challenges associated with the different operation scenarios: high-energy vs low-energy scenarios, high-rates ($Z$ pole) or lower rates ($HZ$ threshold and above).

\subsubsection*{Target methods to be developed}
\begin{itemize}
    \item Strange tagging (see FT {\bfseries HtoSS}, Sec.~\ref{sec:HtoSS}).
    \item Separating lightest families ($u$ and $d$) using final state photon radiation?
    \item Improved $b$ and $c$ jet identification, for example using Machine Learning techniques \cite{JetFlavFCCee,JetFlavILC,JetFlavCepC}.
    \item Improved distinguishing of quark and anti-quarks ($b-\bar{b}, c-\bar{c}, s-\bar{s}$) by displaced vertex charge measurement, charged lepton and kaon identification.
    \item Full $\tau$ lepton reconstruction making use of all detector information (esp. impact parameters, high granularity ECAL for neutral pions); optimal extraction of polarimeter information in many decay modes.
\end{itemize}

\subsection*{MC samples needed}
Basic samples are available as listed in the Motivation Section. Dedicated samples might be needed.

\subsection*{Contact \& Further Information}
\begin{itemize}
\item Gitlab wiki: \url{https://gitlab.in2p3.fr/ecfa-study/ECFA-HiggsTopEW-Factories/-/wikis/FocusTopics/TwoF}
\item Sign up for egroup: ECFA-WHF-FT-TwoF@cern.ch via \url{http://simba3.web.cern.ch/simba3/SelfSubscription.aspx?groupName=ecfa-whf-ft-twof}
\item and/or email the conveners of ECFA WG1 HTE group:\\ \url{mailto:ecfa-whf-wg1-hte-conveners@cern.ch}
\end{itemize}

\clearpage
\section{{\bfseries BCfrag} and {\bfseries Gsplit} --- Heavy quark fragmentation and hadronisation, gluon splitting and quark-gluon separation} 
\label{sec:BCfrag}

\begin{center}
%{\itshape Expert Team: Eli Ben Haim, Loukas Gouskos, Simon Pl\"atzer, Andrzej Siodmok, Torbj\"orn Sj\"ostrand,
%Maria Ubiali, Paolo Azzurri, Ayres Freitas, Adri\'an Irles, Andreas B.\,Meyer}
{\itshape Expert Team: Paolo Azzurri, Eli Ben Haim, Loukas Gouskos, Ayres Freitas, Adri\'an Irles, Andreas B.\,Meyer, Simon Pl\"atzer, Andrzej Siodmok, Torbj\"orn Sj\"ostrand,
Maria Ubiali}
\end{center}

%Meetings of the expert team took place on 31 August and 3 October 2023~\cite{expertteam_bfrag_1,expertteam_bfrag_2}.

%\section{{\bfseries BCfrag} -- Measurement of $b$- and $c$-fragmentation functions and hadronisation rates ($\sqrt{s}=M_Z$ and beyond) (FM)}

Hadronisation and fragmentation of heavy quarks as well as gluon splitting to heavy quarks are key ingredients to the precision modelling of physics processes.
The two issues are entangled, and their systematic uncertainties are expected to be limiting factors in precision Higgs and electroweak measurements at the Higgs factory. To assess their impact, it is important to review the state of the art, to project how future theory developments could reduce current uncertainties, and to devise ways to constrain heavy quark fragmentation and gluon splitting from measurements at the Higgs factories. 

In the high-precision limit, fragmentation functions will not be universal, i.e.\ they are expected to depend on observables and initial states. It is argued that the factorisation of the perturbative and non-perturbative parts of the problem is not possible without dedicated tuning of free parameters in the required fragmentation model used. There are ongoing developments in disentangling hadronisation and fragmentation, specifically the cross-talk between parton shower and fragmentation. New work is needed for NLL-accurate showers.

The splitting of gluons into $b\bar{b}$ or $c\bar{c}$ is only modelled in the perturbative step of the process (not in the string/cluster fragmentation) but still, charm and bottom masses are parameters in the shower. Other relevant differences may be in the treatment of the strong coupling differently for heavy quarks or (massless) gluons in the limit of $p_{\rm T}=0$. Separation of $h\to$ gluons from $h \to b\bar{b}/c\bar{c}$ ($\sqrt{s}=M_Z$ and beyond) is affected.
Details are described in Ref.~\cite{sjostrand_gQQ}.

There has been a lot of progress recently on the computation of perturbative bottom and charm fragmentation functions (FFs).  Heavy quark FFs can be computed
in perturbation theory in QCD, starting from initial conditions at a reference scale $\mu_0\sim m_Q$ (with $m_Q$ the mass of the heavy quark) 
and employing the timelike DGLAP evolution equations to evolve them up to any other scale. 
Initial conditions for the gluon- and heavy-quark-initiated fragmentation into a
heavy quark are known at order $\alpha_{\rm S}$~\cite{Mele:1990cw,Mele:1990yq} and have been computed at order $\alpha_{\rm S}^2$\cite{Melnikov:2004bm,Mitov:2004du}. 
The timelike DGLAP evolution equation is implemented in public codes such as {\sc QCDnum}~\cite{Botje:2010ay},
{\sc ffevol}~\cite{Hirai:2011si}, {\sc APFEL}~\cite{Bertone:2013vaa} or {\sc MELA}~\cite{Bertone:2015cwa}, up to NNLL logarithmic accuracy.  
In Ref.~\cite{Ridolfi:2019bch}, the role of gluon-initiated fragmentation to heavy quarks has been considered, and the coupled timelike 
evolution of bottom quarks and gluons is considered in detail. This is important as, while at LEP and at Tevatron the 
$g\to b\bar{b}$ splitting mechanism was considered subdominant, this is no longer the case at
the LHC. In Ref.~\cite{Maltoni:2022bpy} the perturbative component of the fragmentation function of
the $b$ quark to the best of the present theoretical knowledge was presented. The fixed-order calculation
to order $\alpha_{\rm S}^2$ of the fragmentation function at the initial scale~\cite{Melnikov:2004bm,Mitov:2004du} is matched with soft-emission
logarithm resummation to next-to-next-to-leading logarithmic accuracy, so that order-$\alpha_{\rm S}^2$
corrections are accounted for exactly, and logarithmically enhanced contributions from
loops of $b$ quarks are included. In Ref.~\cite{Czakon:2021ohs} the perturbative computation of the $b$-quark fragmentation function at NNLO + NNLL is supplemented 
by the fit of the non-perturbative component $b\to B$ hadrons. Similarly, the perturbative component of $c$-quark fragmentation and the fit of the non-perturbative component $c\to D$ hadrons are discussed in Ref.~\cite{Bonino:2023vuz}. A key question is how can this progress in the Perturbative FF framework, supplemented by fits of the non-perturbative component, could be implemented in practice, for example in PS Monte Carlo.

The perturbative step primarily describes the $b/c$ quark inclusive fragmentation function. The step from there to $B/D$ hadrons is of
non-perturbative character. As noted above, inclusive non-perturbative fits can be used to supplement the perturbative calculations.
The exclusive nature of the fragmentation process is not addressed in such approaches, however. Concretely, the modifications under
experimental cuts and selection criteria is undefined. Therefore parton showers are needed for the perturbative step, and fragmentation
models for the non-perturbative one. These two steps are intertwined and convoluted in a complicated space of momenta and colours, so that fragmentation in general cannot be described as a convolution of a
perturbative state and a fragmentation function $f(z)$ with $0 < z < 1$.
%Fragmentation cannot be described as a convolution of a perturbative matrix element and a fragmentation function $f(z)$ with $0<z<1$. 
In string fragmentation, the colour strings stretched between partons can be attached so as to pull a heavy hadron either to
a lower or to a higher momentum than the mother quark, depending on the string topology at hand \cite{Norrbin:2000zc}. Thereby
also $z > 1$ becomes possible, which for some observables can have a significant impact relative to naive expectations. %In string fragmentation, colour strings stretching between partons cause acceleration and produce a {\bf\boldmath hardening} of the hadron spectra at high energies~\cite{sjostrand_fwdcharm}.
Similar considerations also exist for cluster fragmentation models.

\subsection{Relevance for the physics program of a Higgs/Top/EW Factory}
Jets and in particular heavy-flavour jets play an important role in many of the flagship measurements of Higgs/Top/EW Factories. As examples, we highlight here the connection to other ECFA Focus Topics:

\begin{description}

\item[\boldmath Precise study of $h\rightarrow gg/b\bar{b}/c\bar{c}$: ] {\bf (HtoSS, Sec.~\ref{sec:HtoSS}, and ZHang, Sec~\ref{sec:ZHang})}.
Future Higgs Factories will provide sensitivity to these topologies providing capabilities to fully explore the second generation of Yukawa couplings, which is out of reach at the LHC. However, current uncertainties in gluon splitting into heavy quarks would introduce large systematic uncertainties in the measurements. The questions arising are: how to consistently implement gluon splitting in parton shower tools (modelling and free parameters)? and how to evaluate the impact of incomplete modelling of the gluon splitting when determining the $h\rightarrow gg/b\bar{b}/c\bar{c}$  couplings? This issue is discussed in the \textbf{HtoSS} focus topic (Sec.~\ref{sec:HtoSS}).

\item[\boldmath Precise determination of $W$-mass and cross section:] {\bf (Wmass, Sec.~\ref{sec:Wmass}, Wdiff, Sec.~\ref{sec:WWdiff}, and CKMWW, Sec.~\ref{sec:CKMWW})}.  
$W$ mass measurements at future Higgs factories are expected to deliver statistical accuracies at the MeV level  \cite{LCCPhysicsWorkingGroup:2019fvj, Azzurri:2021yvl}. To match this unprecedented precision, the control of systematic uncertainties is crucial. At LEP2, the modelling of non-perturbative QCD effects in $W$ boson
hadronic decays was a dominant source of systematic uncertainties. Further theoretical and experimental studies are required to estimate the size of such uncertainties at future colliders.
This issue is discussed in the \textbf{Wmass} focus topic.

\item[\boldmath $Z$-$b/c$ couplings:] {\bf  (TwoF, Sec.~\ref{sec:TwoF})}.
What would be the impact of these uncertainties on the extraction of $Z$-$b/c$ couplings at the $Z$-pole?  In Ref.~\cite{AlcarazMaestre:2020fmp} it is demonstrated that hadronisation uncertainties have a significant impact on determinations of the partial widths normalised to total hadronic width ($R_{b,c}$), the forward-backward asymmetries ($A_{\rm FB}^{b,c}$)  (or left-right asymmetries) at $e^+e^-$ colliders, even after application of cuts to reduce their impact. The size of these uncertainties could be a limiting factor when operating at the $Z$-pole in the high luminosity scenarios of FCC-ee. 
This issue is discussed in the \textbf{TwoF} focus topic.

\end{description}

\subsection{Theoretical/phenomenological targets}

\begin{itemize}
    \item Are existing hadronisation models (strong fragmentation, cluster fragmentation) flexible enough, or do we need new ideas like for example ML hadronisation models~\cite{Ghosh:2022zdz,Ilten:2022jfm,Chan:2023ume,Chan:2023icm}? Can perturbatively constrained models \cite{Christiansen:2015yqa,Gieseke:2017clv,Gieseke:2018gff,Platzer:2022jny} reduce uncertainties?
    \item Identify calibration observables well understood theoretically and unaffected by BSM physics.
    \item Treatment of photons and gluon splitting.
    \item See also Ref.~\cite{ecfa22_kraus}
\end{itemize}

\subsection{Target physics observables}

A summary of recommended measurements is given in 
Table~\ref{table:bcfrag}. Related observables for $ee$ and $pp$ are compared.
In practice, the measurements of the observables would be intertwined. 
The following measurements should also be considered:

% could make this into a table.
\renewcommand{\arraystretch}{1.3}
\begin{table}
\caption{Target physics observables at $e^+e^-$ and $pp$.}
\rotatebox{-90}{
\small
\begin{tabular}{|p{0.30\textheight}p{0.25\textheight}p{0.35\textheight}|}
\hline
{\bf\boldmath Observable} & $e^+e^-$ & $pp$ \\
\hline
{\bf\boldmath Event shapes and angular distributions} & & \\
{\bf\boldmath Inclusive $B/D$ production cross section} & 
primary production is well known from theory, so any "excess" is from gluon splitting &
combines primary production, gluon splitting, and MPI (multiparton interactions) contributions, each with significant theoretical uncertainties \\
{\bf\boldmath Flavour composition} as far back in decay chains as can be traced (even equal $D^{*0}$ and $D^{*+}$ rates gives unequal $D^0$ and $D^+$ ones) &
we do not expect sizeable momentum dependence, but interesting to contrast mesons and baryons for smaller ones &
significant $p_{\rm T}$ dependence observed and to be studied further, also high- vs. low-multiplicity events, rapidity, ..., which is important for development/tuning of colour reconnection models \\
{\bf\boldmath Particle-antiparticle production asymmetries} &
none expected, except tiny from CP-violation in oscillations &
asymmetries expected and observed from $p$ flavour content, increasing at larger rapidities; relates to how string (and cluster?) fragmentation connects central rapidities to beam remnants \\
{\bf\boldmath Momentum spectra} &
$dn/dx_E$ with \mbox{$x_E$ = $2E_{\rm had} / E_{\rm cm}$}; basic distribution for tuning of "fragmentation function" &
$dn/dp_{\rm T}$ and $dn/dy$ give basic production kinematics, but the many production channels give less easy interpretation \\
{\bf\boldmath Energy flow around $B/D$ hadrons}, excluding the hadron itself, as a test that dead cone effects are correctly described &
$dE/d\theta$ where $\theta$ is the distance from $B/D$ on the sphere &
$dp_{\rm T}/dR$ where $R$ is the distance in ($\eta$, $\phi$) or ($y$, $\phi$) space, only applied for $B/D$ above some $p_{\rm T}$ threshold \\
{\bf\boldmath $B/D$ hadron momentum fraction} of total $E$ or $p_{\rm T}$ in a jet, with $x = p_{\rm T}^{\rm had} / p_{\rm T}^{\rm jet}$, as a test of the fragmentation function combined with almost collinear radiation, suitably for some slices of $p_{\rm T}$ (and in addition with a veto that no other $B/D$ should be inside the jet cone, so as to suppress the gluon splitting contribution) &
draw a jet cone in $\theta$ around $B/D$ and measure $x$ &
draw a jet cone in $R$ around $B/D$ and measure $x$ \\
{\bf\boldmath $B/D$ hadron multiplicity}, as a measure of how often several pairs are produced & & \\
{\bf\boldmath Separation inside $B/D$ pairs}, where large separation suggests back-to-back primary production, while small separation suggests gluon splitting &
separation in $\theta$ &
separation both in $\phi$ and in $R$, since for primary production $\phi = \pi$ is hallmark with $\eta/y$ separation less interesting, while gluon splitting means $R$ is small while $\phi$ and $y/\eta$ individually are less interesting \\
{\bf\boldmath Hardness difference} within (reasonably hard) pairs, $\Delta = (p_{\rm T}^{\rm max} - p_{\rm T}^{\rm min}) / (p_{\rm T}^{\rm max} + p_{\rm T}^{\rm min})$, where for gluon splitting $x^2 + (1 - x)^2$ translates to $1 + \Delta^2$ &
separately for small or large $\theta$ &
separately for large or small $\phi$ \\[0.5mm]
\hline
\end{tabular}
}
\label{table:bcfrag}
\end{table}

\begin{itemize}
\item For a pair with small separation, say $\theta/R < 0.7$, draw a cone around the midpoint of the two, say again $\theta/R = 0.7$, and find the fraction $x = (p_{\rm T}^{{\rm had},1} + p_{\rm T}^{{\rm had},2}) / p_{\rm T}^{\rm jet}$, to quantify loss to showers and hadronisation. This loss would be reduced by colour reconnection which could combine the $b\bar{b}$ or $c\bar{c}$ quark pairs into a singlet, rather than the default octet where the two pairs fragment separately.
\item In events with two $B/D$ pairs, many observables become possible. There are four possible particle-antiparticle pairs (more if $B$ mixing is considered), each of which can be studied according to the two points above. In addition, a pair with a small separation would suggest a gluon splitting, while one with a large ditto is a primary production. For $pp$, two back-to-back pairs would suggest MPI. One can try to classify events into most likely history and study the relative composition of (a) two separate hard processes (MPIs, $pp$ only); (b) one hard process and one gluon split; (c) two gluon splits on the same side of the event; and (d) two gluon splits on opposite sides.
\item Even if one $B/D$ is missed in $pp$ collisions, and only three $B/D$ hadrons are observed, one can study the three pairings and see whether either pair has a small $R$ or a large $\phi$. Again relative rates will provide info on the composition of production mechanisms.
\end{itemize}

\subsection{Target detector performance, analysis methods and tools}
\begin{itemize}
    \item Large tracker acceptance as well as very good vertexing and flavour tagging capabilities (including light quarks and gluon quarks)
    \item Jet charge measurements, including charge hadron identification capabilities (see above), and fit a representative set of observables for hadronisation calibration (see above).
    \item Samples for hadronic observables using different hadronisation models and parameters. Full simulation is required to understand flavour tagging capabilities. Existing tools are e.g. the generators \textsc{Pythia}~\cite{Bierlich:2022pfr}, \textsc{Herwig}~\cite{Bahr:2008pv,Bellm:2015jjp}, \textsc{Sherpa}~\cite{Sherpa:2019gpd} and the tuning tools \textsc{Professor}~\cite{Buckley:2009bj}, \textsc{Rivet}~\cite{Bierlich:2019rhm}.
    \item Access to LEP Archived Data. LEP data (and simulations) have been partially archived to allow their use for physics analyses after the closure of the collaboration. The use of archived data is authorised to former members of the collaboration and collaborators. However, the understanding and reprocessing of analysis is still challenging and depends on the safeguarding of the different collaboration’s analysis frameworks and mini-data at CERN. Recent efforts \cite{Badea:2019vey} have been driven to re-analyse these data by exporting the data and simulations to more modern and accessible formats, for instance, the MIT Open Data format. A systematic approach for the exportation of such archived data and software tools to the {\sc Key4hep} environment should be considered by the Higgs Factory community. This would allow the validation of newer calculations and MC tools with existing data.
\end{itemize}    

\subsection{Summary and open questions}
\begin{enumerate}
    \item How can the recent progress in the Perturbative FF framework, supplemented by the fits of the non-perturbative component, be implemented and used in practice, e.g.\ in PS Monte Carlo?
    \item What is the quantitative impact of uncertainties from parton-shower, fragmentation and hadronisation on flagship Higgs/Top/EW measurements - and which level of precision will be required? 
    \item Which measurements of particle rates, species, distributions are needed in order to constrain fragmentation and hadronisation models to the required level of precision?
    \item Which detector capabilities are required and to which extend do the proposed detector concepts provide these?
    \item To which extend could LEP data be useful and how could they be made accessible to test new calculations and MC tools? 
\end{enumerate}

\subsection{Contact \& Further Information}
\begin{itemize}
\item Gitlab wiki: \url{https://gitlab.in2p3.fr/ecfa-study/ECFA-HiggsTopEW-Factories/-/wikis/FocusTopics/BCFRAG}
\item Sign up for egroup: ECFA-WHF-FT-BCFRAG@cern.ch via \url{http://simba3.web.cern.ch/simba3/SelfSubscription.aspx?groupName=ecfa-whf-ft-bcfrag}
\item and/or email the conveners of ECFA WG1 PRECision group:\\ \url{mailto:ecfa-whf-wg1-prec-conveners@cern.ch}
\end{itemize}

\clearpage
%\section{{\bfseries Gsplit} -- Measurement of gluon splitting to $bb$ / $cc$, interplay with separating $h\to$ gluons from $h \to bb / cc$ ($\sqrt{s}=M_Z$ and beyond) }

%\begin{center}
%{\itshape
%Expert Team:
%Eli Beh-Haim, 
%Loukas Gouskos, 
%Simon Plaetzer, 
%Andrzej Siodmok, 
%orbjorn Sjostrand, 
%Maria Ubiali,\dots
%}
%\end{center}

%\input{sections/gsplit}
%\clearpage

%%%%%%%%%%%%%%%%%%%%%%%%%%%%%%%%%%%%
\addcontentsline{toc}{section}{References}
%\setboolean{inbibliography}{true}
\bibliographystyle{LHCb}
\bibliography{sections/bcfrag,sections/exscalar,sections/hself,sections/lumi,sections/wdiff,sections/bktautau,sections/extt,sections/htoss,sections/ttdet,sections/wmass,sections/ckmww,sections/llps,sections/twof,sections/zhang,benchmarks}

%%%%%%%%%%%%%%%%%%%%%%%%%%%%%%%%%%%%
\end{document}